\documentclass[%
reprint,
 amsmath,amssymb,
 aps,pre
]{revtex4-1}

\usepackage{amsmath}
\usepackage{amssymb}
\usepackage{float}
\usepackage{graphicx}
\usepackage{physics}
\usepackage{txfonts}
\usepackage{hyperref}

\usepackage{soul}

\newcommand{\fig}{Fig.~}
\newcommand{\eq}{Eq.~}
\newcommand{\rb}{{\mathbf r} }
\newcommand{\um}{\muup\textrm{m}}
\graphicspath{{img/}}

\begin{document}

\title{Sensing the shape of a cell with reaction-diffusion and energy minimization}

\author{Amit R. Singh}
\affiliation{Department of Mechanical Engineering, Birla Institute of Technology and Science, Pilani}
\author{Travis Leadbetter}
\affiliation{Department of Physics \& Astronomy, Johns Hopkins University, Baltimore, MD.}
\author{Brian A. Camley}
\affiliation{Department of Physics \& Astronomy, Johns Hopkins University, Baltimore, MD.}
\affiliation{Department of Biophysics, Johns Hopkins University, Baltimore, MD.}

\begin{abstract}

 Some dividing cells sense their shape by becoming polarized along their long
 axis. Cell polarity is controlled in part by polarity proteins like Rho GTPases
 cycling between active membrane-bound forms and inactive cytosolic forms,
 modeled as a ``wave-pinning'' reaction-diffusion process. Does shape sensing
 emerge from wave-pinning?  We show that wave pinning senses the cell's long
 axis. Simulating wave-pinning on a curved surface, we find that high-activity
 domains migrate to peaks and troughs of the surface. For smooth surfaces, a
 simple rule of minimizing the domain perimeter while keeping its area fixed
 predicts the final position of the domain and its shape.  However, when we
 introduce roughness to our surfaces, shape sensing can be disrupted, and
 high-activity domains can become localized to locations other than the global
 peaks and valleys of the surface. On rough surfaces, the domains of the
 wave-pinning model are more robust in finding the peaks and troughs than the minimization rule, though both can become trapped in steady states away from the peaks and valleys. We can control the robustness of shape sensing by altering the Rho GTPase diffusivity and the domain size. {We also find that the shape sensing properties of cell polarity models can explain how domains localize to curved regions of deformed cells.}
 Our results help to understand the factors that
 allow cells to sense their shape – and the limits that membrane roughness can
 place on this process.

\end{abstract}

\maketitle
For cells to respond to changing environments~\cite{doyle_2013_dimensions,baker_2012_deconstructing} 
by choosing a direction to crawl, an axis of division, or a
location to form a new branch, they must develop an internal biochemical
polarity, where proteins are distributed inhomogeneously around the cell surface. In
addition, the shape of the cell and its internal membranes can help organize its
polarity
-- cells can sense their own shape \cite{haupt_2018_how}. Localization of different proteins 
to different regions can occur when individual proteins prefer to bind to membranes that have a specified curvature range, as is known to happen with BAR proteins
~\cite{simunovic_2016_how,peter_2004_bar},
ArfGAP~\cite{bigay_2005_arfgap1,drin_2007_general,vanni_2014_sub},
\(\alpha\)-synuclein ~\cite{pranke_2011_alpha}
SpoVM~\cite{levin_1993_unusually} and septins~\cite{mostowy_2012_septins}.
However, even if individual proteins do not have a curvature preference,
reactions on a cell membrane can be sensitive to the shape of that membrane, {leading to patterns of protein localization sensitive to the membrane's shape. This is the broadest idea of ``shape sensing."}
This shape sensitivity can arise from the local changes in surface-to-volume
ratio \cite{meyers_2006_potential} or more complicated reaction-diffusion
mechanisms
\cite{camley_2017_crawling,cusseddu_2019_coupled,rangamani_2013_decoding,ramirez2015dendritic,frank2019pinning,spill2016effects,eroume2021influence}.
Experiments have measured key correlations between localization of myosin II
\cite{elliott_2015_myosin}, and PIP$_2$ \cite{driscoll_2019_robust} and local
cell shape features like curvature, while inducing curvature within a cell can
change the localization of both myosin II \cite{elliott_2015_myosin} and the
polarity protein Rho \cite{wigbers_2021_hierarchy}. {Yeast polarity domain size also reflects cell shape \cite{bonazzi2015actin}.} Bacterial shape sensing has also been
recapitulated {\it in vitro} for the Min system of E. coli
\cite{schweizer_2012_geometry,halatek_2014_effective}. Shape sensing may also
play a role in creating instabilities where cells crawl in circles
\cite{camley_2017_crawling,allen_2020_cell} or have a periodic motion
\cite{camley_2013_periodic}. Even the existence of polarization is sensitive to
cell volume \cite{hubatsch_2019_cell}. Most dramatically, recent work studying
the distribution of PAR proteins in the \textit{C. elegans} zygote demonstrates
a clear binary shape sensing: when PAR proteins are disrupted from their
natural location by experimental intervention, they return to become localized
to the {\it nearest} narrow end of the zygote ~\cite{mittasch_2018_non}. {This experiment
demonstrates that PAR proteins sense the long axis of the cell  -- while not establishing whether this sensing is self-organized, or arises from a pre-existing pattern.}

{Some elements of shape sensing are well understood, e.g. how patterns can be selected by controlling the possible wavelengths of an initial linear instability \cite{murray2001mathematical}. 
Here we focus on a specific aspect of shape sensing: motivated by \cite{mittasch_2018_non}, we study how a single initial domain of high concentration moves in response to the shape of its membrane. We will refer to this as ``domain migration shape sensing," to distinguish it from other examples of sensitivity to shape mentioned earlier. Domain migration cannot be captured by linear stability analysis, and is not well understood. As a prototype model, we study a minimal model of cell polarity, the ``wave-pinning" (WP) model of Mori et al.
\cite{mori_2008_wave} which describes Rho GTPase dynamics. We use this model as the simplest model that robustly describes cell polarity, but it is also closely related to more detailed models used to describe PAR protein dynamics \cite{goehring2011polarization}. Previous work has shown that  membrane-bound active form
of Rho GTPases localize to the narrow end of the cell
\cite{jilkine_2009_wave,vanderlei_2011_computational,camley_2017_crawling}, and links between the wave pinning model and the Allen-Cahn model have been suggested \cite{jilkine_2009_wave,mori2011asymptotic}. These behaviors occur not only in the basic wave pinning model but also in significantly more complicated models including multiple Rho GTPases and phosphoinositides \cite{maree2012cells,maree2006polarization}.
Later work has also shown that narrow-end localization occurs in three dimensions, but also demonstrates that domain localization in response to a complex cell shape is difficult to predict \cite{cusseddu_2019_coupled}.}

{Can we reproduce the  shape
sensing via domain migration of \cite{mittasch_2018_non} using a simple model?  Is there a predictive minimal theory for
where polarity proteins will end up in a cell, other than just solving the
complex reaction-diffusion PDEs? Previous simulation work has also focused on
smooth, idealized surfaces. Would reaction-diffusion mechanisms for shape
sensing be robust to the rough, complex shapes observed in real cells? In this
paper, we study these questions. We find that a minimal WP model is able to
recapitulate binary shape sensing as well as localization of domains to corners of triangular cells. }We also argue that in many cases, we can
understand the dynamics of complex domain shapes arising from the WP model by
minimizing domain perimeter while keeping domain area fixed. However, we find
that this simple minimization principle can be disrupted in sufficiently rough
membrane geometries, where shape sensing itself is also less reliable. We show
that shape sensing can be modulated by altering the diffusion coefficient of
the membrane-bound form of our Rho GTPase, as well as the domain area. Our work is a
systematic test of a minimal theory for how shape influences
Rho GTPase cell polarity, and provides an understanding of when shape sensing will succeed
depending on the cell geometry. 

\begin{figure}[htb]	
	\includegraphics{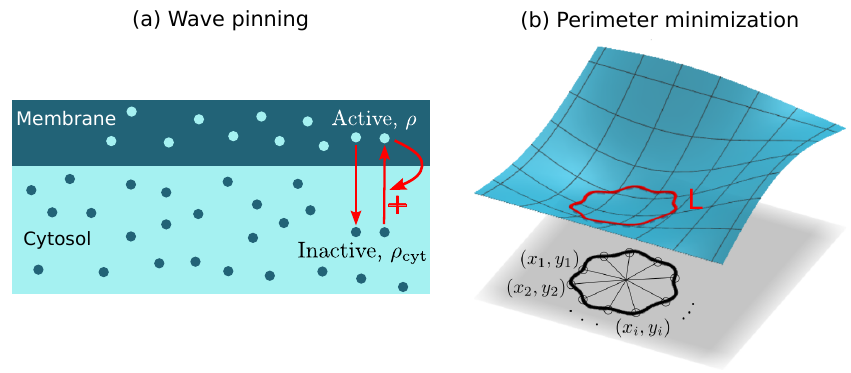}
	\caption{{\textbf{(a)} A schematic of the reaction
	involved in the wave-pinning (WP) model. In the inactive form, Rho GTPases are in the
	cytosol, $\rho_\textrm{cytosolic}$. The active form $\rho$ is in the membrane. 
	There is a positive feedback: presence of the active form on the membrane locally promotes 
	conversion from the inactive form.
	\textbf{(b)} Representation of a domain in the perimeter minimization (PM) model. The red curve lies in a 3D surface. Its projection on the
	\(x\)-\(y\) plane is shown as the black curve. To calculate the length of the
	boundary and the surface area enclosed by the red curve, we divide the black
	curve into triangles as shown and use results from differential geometry (Appendix).}}
 \label{fig:bothmodels}
\end{figure}

\section{Models}

\subsection{Wave-pinning model of cell polarity}

We describe cell polarity with a variant of the wave-pinning (WP)
reaction-diffusion system~\cite{mori_2008_wave}. This model treats Rho GTPases
exchanging between the active membrane-bound $\rho(\rb)$ and inactive cytosolic
$\rho_\textrm{cytosolic}$ states with rate $f(\rho,\rho_\textrm{cytosolic})$,
and the membrane-bound form diffusing on the curved membrane with diffusion
coefficient $D$; the total amount of Rho GTPase is conserved. $\rho(\rb)$
obeys the reaction-diffusion equation:
\begin{align}
 \label{eq:rho}
 \pdv{\rho}{t} &= D\laplacian \rho + f(\rho, \rho_{\mathrm{cytosolic}}).
\end{align}
The reaction term \(f(\rho, \rho_{\mathrm{cytosolic}})\) includes basal rates
of activation, de-activation, and a positive feedback where activation occurs
more often when active Rho GTPase is already present:
\begin{align}
	\label{eq:reaction}
 f(\rho, \rho_{\mathrm{cytosolic}}) = \rho_{\mathrm{cytosolic}}\left(k_0 +
 \frac{\gamma \rho^2}{K^2 + \rho^2}\right) - \delta \rho
\end{align}
where \(k_0\) is a basal rate of activation, \(K\) is the concentration at
which the positive feedback begins to saturate, and \(\gamma\) is the maximal
rate of activation from positive feedback. The basal rate of the reverse
reaction i.e. conversion from \(\rho\) to \(\rho_{\mathrm{cytosolic}}\) is
\(\delta\). In all the results presented in this paper, we have used \(k_0 =
0.07\) s\(^{-1}\), \(\gamma = 5 \) s\(^{-1}\), \(K = 2 \) \(\mu\mathrm{m}^{-2}\), and
\(\delta = 3 \) s\(^{-1}\). Unless otherwise stated, we use \( D = 0.5\
\mu\mathrm{m}^{2}\mathrm{s}^{-1}\). The order of magnitude of these quantities
are based on \cite{mori_2008_wave}. 

We assume that the diffusion coefficient of the cytosolic form is so much
larger than the diffusion coefficient of \(\rho\) that
\(\rho_{\mathrm{cytosolic}}\) is well-mixed, i.e. constant over the cell
volume. {Because of the conservation of Rho GTPases between the membrane-bound
and cytosolic forms, the total number of Rho GTPase proteins is $\int_\textrm{membrane} \rho + \int_\textrm{cytosol}
\rho_\textrm{cytosol}$, which is a unitless constant we call \(M\)}. We can then determine
$\rho_\textrm{cytosolic}$ as:
\begin{align}
 \label{eq:rhocyt}
 \rho_{\mathrm{cytosolic}} &= \frac{M}{\omega S} - \frac{1}{\omega S}\int_{\mathrm{membrane}} \rho.
\end{align}
\(S\) denotes the surface area of the cell. \(\omega\) is the ratio of the
volume of the cell to its surface area. Thus, \(\omega S\) is the volume of the
cell. When we simulate an abstract surface \(h(x, y)\) where there is no clear
cytosolic volume, we take \(\omega=1\,\um\). We choose \(M/S = 2.9 \mu m^{-2}\)
as a default value in our simulations; changing this value changes the size of
the domain of high Rho GTPase activity, and if it is increased or decreased too
much, it will prevent the cell from polarizing.  The equation \eq \ref{eq:rho} is
solved on our curved surfaces using a finite element method (Appendix). 

\begin{figure*}[htb]
	\centering
	\includegraphics{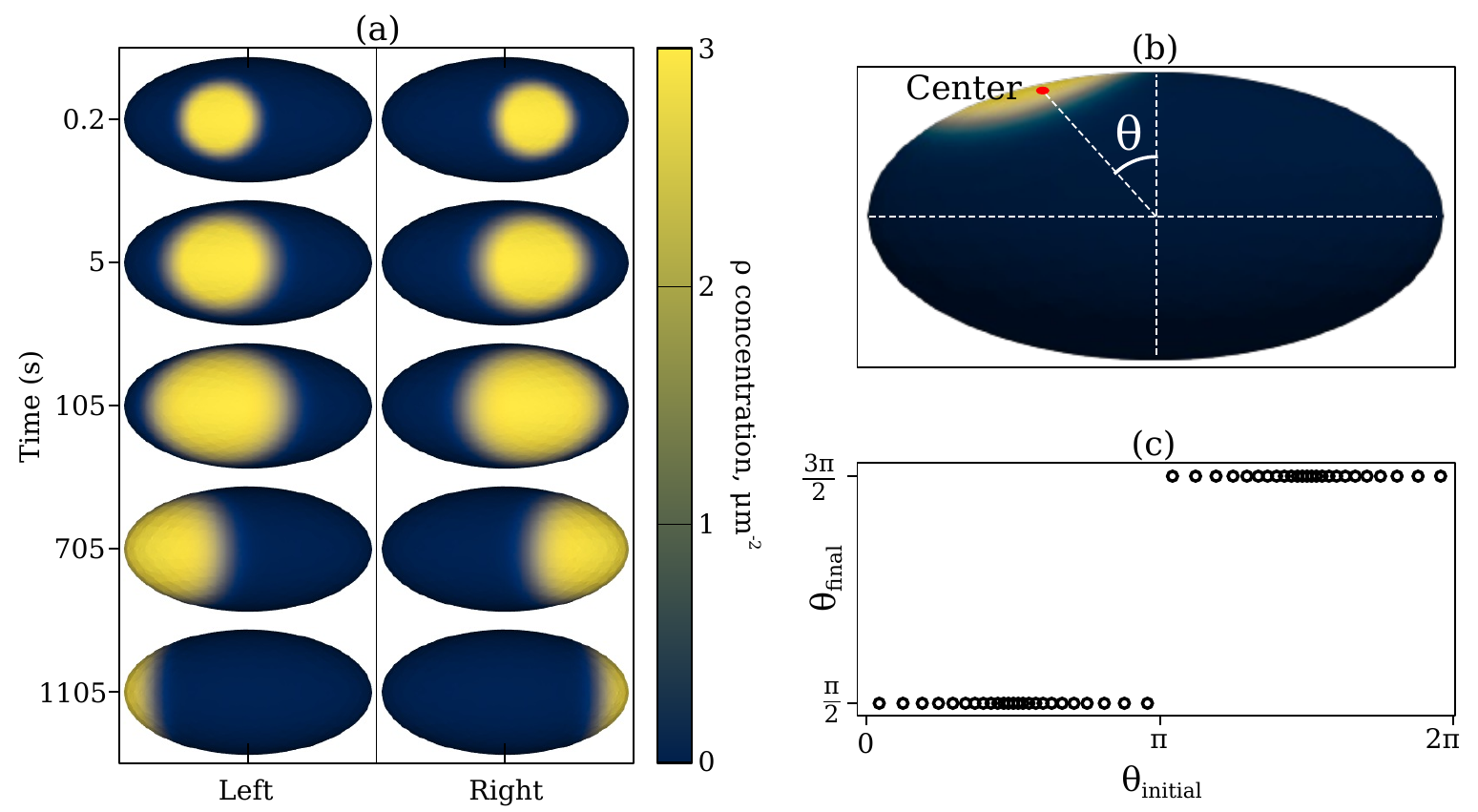}

	\caption{The WP model is able to reproduce the shape sensing behavior of PAR
	proteins. {\bf (a)} Top view: High concentration domains that were created
	away from the ends travel towards the ends over a duration of several
	minutes. {Note that we initialize the domain at a size a little smaller than its steady-state size; it first expands and then migrates.} {\bf (b)} Front view: the angular position of the center of mass
	of a high-concentration domain from the vertical axis. {\bf (c)} the final
	angular position of the domain center of mass as a function of the inital angular
	position. The steady state ``flips'' when $\theta$ crosses $\pi$. This is analogous to the binary shape sensing of \cite{mittasch_2018_non}.}

\label{fig:ellipsoid}
\end{figure*}

When parameters allow
polarization (see, e.g.
\cite{mori_2008_wave,camley_2013_periodic,mori2011asymptotic}), the WP model
reaction-diffusion equation \eq \ref{eq:rho} admits solutions that have a
high-concentration region with $\rho \approx \rho^+$ and low-concentration
regions of $\rho \approx \rho_-$. These values are set by the roots of $f(\rho,\rho_\textrm{cytosolic})=0$, which are, when the system allows polarization, \(\rho_-, \rho_0\), and
\(\rho_+\), in increasing order.  {An initial domain of enriched $\rho$ will evolve into 
a ``pinned" state, where it has an area set by the reaction kinetics, the geometry, and the total amount of Rho GTPase $M$. The conditions for pinning, which set the steady-state domain shape in terms of the total amount $M$ and the values \(\rho_-\) and \(\rho_+\) are discussed in \cite{mori_2008_wave,mori2011asymptotic,camley_2013_periodic}.}
After the formation of a stable pinned domain, the domain moves over
the surface, keeping its area roughly constant. 
On a flat two-dimensional surface, Jilkine has used asymptotic
analysis to show that the velocity of the domain edge is is proportional to its
curvature~\cite{jilkine_2009_wave}, tending to minimize the perimeter.

\subsection{Perimeter minimization model}

On a flat surface, solutions of the wave-pinning equations tend
to minimize their domain perimeter \cite{jilkine_2009_wave}, but their area
remains roughly constant, suggesting that the dynamics of domains in these
reaction-diffusion systems may be captured by minimizing perimeter while keeping
area constant. This same idea may also capture more general proposed mechanisms
for cell polarity, like binary mixture phase
separation~\cite{orlandini_2013_domain,semplice2012bistable}. In these models,
line tension between the two phases drives domain coalescence and coarsening to
minimize interface tension, while mass conservation keeps domain area fixed.
Thus, both wave-pinning and phase separation pictures of cellular polarization
tend to minimize the interface length of a polarity protein domain. Therefore,
we will also study a perimeter minimization (PM) model of cell polarity. In the perimeter minimization
 model, the high \(\rho\) concentration domain is represented by a portion
of the 3D surface bounded by a closed curve that lies in the surface (Fig. \ref{fig:bothmodels}b). The closed
curve is the boundary of the domain. We minimize the length of the boundary
while constraining the enclosed surface area to be fixed. The domain is free to
relocate on the underlying surface. 

To calculate the length of the boundary and the enclosed area, we project the
3D curve to the \(x\)-\(y\) plane, as shown in \fig\ref{fig:bothmodels}.
The projected curve is divided into triangular ``pie slices''.
Given the surface $h(x,y)$ and the points $\{x_i,y_i\}$
defining the domain {we can compute the domain's
perimeter $L\left(\{x_i,y_i\}\right)$ and area $A\left(\{x_i,y_i\}\right)$ on the surface using the first fundamental form of the surface and Gaussian quadrature (see
Appendix). We chose this approach to allow for simple differentiation of the energy.} We have argued that we should expect many polarity mechanisms to
minimize the domain perimeter while keeping domain area fixed. To numerically
find these minima, we minimize the energy
\begin{align}
 \label{eq:perimin}
 \mathcal{F}\left(\{x_i, y_i\}\right) = L\left(\{x_i, y_i\}\right) +
 \frac{1}{2}k\left(A\left(\{x_i, y_i\}\right) - A_0\right)^2
\end{align}
Eq. \ref{eq:perimin} penalizes deviations away from the prescribed area $A_0$
with a coefficient $k = 10^3 \mu\mathrm{m}^{-1}$ as a strong area constraint. (We find that with this value of $k$, steady-state areas are well-constrained to $A_0$ to within about 2\%).  {We choose \(A_0\) to be the average steady state area of high
concentration domains in WP simulations for the given surface. The parameter $A_0$ will also depend on the total amount of 
protein $M$, because this will alter the steady-state domain size. Values of $A_0$ for each simulation are provided in Table \ref{table:param} in the Appendix.}

We assume that the domain evolves in an overdamped way, i.e. the velocity of a point $(x_i,y_i)$ is negatively proportional to the gradient of the energy $\mathcal{F}$. {The overdamped dynamics assumption here is a minimal one; more complex models that still minimize the energy would also be possible \cite{hohenberg1977theory}, e.g. modeling how the membrane lipids flow in response to deformations of a domain with a line tension \cite{camley2010dynamic}.} The overdamped dynamics corresponds to minimizing $\mathcal{F}$ using a simple gradient-descent
algorithm. 
Thus, we generate a series of \(\{x_i, y_i\}\) that converge to a local minimum.
The update from the \(n^{\mathrm{th}}\) iteration to the \((n +
1)^{\mathrm{th}}\) iteration is obtained as 
\begin{align}\label{eq:gradient_descent}
\begin{split}
x^{n+1}_i &= x^{n}_i - \beta \pdv{\mathcal{F}}{x^{n}_i}\\
y^{n+1}_i &= y^{n}_i - \beta \pdv{\mathcal{F}}{y^{n}_i}
\end{split}
\end{align} where \(\beta\)
controls the step-size along the gradient. To ensure steady state, we continue
to evolve the system until the solution remains unchanged {to a
precision of \(10^{-6}\)} for 1000 iterations. {We use
\(\beta = 10^{-2}\mu\mathrm{m}\).}

\section{Results}
\label{sec:results}

\subsection{Wave-pinning exhibits binary shape sensing}
\label{subsec:bistability}

One of the motivating experimental results for this study is the reorientation
of PAR protein domains in a \textit{C. elegans} zygote~\cite{mittasch_2018_non}
to regions of high curvature. {In these experiments, a \textit{C. elegans} zygote
of roughly ellipsoidal shape is subjected to cytoplasmic flows such that the
high partitioning-defective protein (PAR) concentration domain is rotated from
the initial position on one end of the zygote. If the induced rotation is less
than 90\(^{\circ}\), the PAR domain goes back to its initial position. But for
rotations greater than 90\(^{\circ}\) the polarity of the cell is reversed.
Thus, in the steady state, the PAR domain is always on one of the ends of the
long axis of the zygote. {PAR proteins have reaction kinetics similar to the Rho
GTPases i.e. there are membrane-bound active and cytosolic inactive forms that exchange~\cite{hoege_2013_principles}, and PAR models have been built from extending WP models \cite{goehring_2011_polarization}.} Therefore, the WP model 
may be an effective minimal model for some elements of PAR polarity. We test if the WP model
can reproduce this bistability.}  We solve the reaction-diffusion system on an ellipsoidal surface,
with initial conditions describing a high-$\rho$ domain at different angles
$\theta$ (\fig\ref{fig:ellipsoid}b). The WP model is able to reproduce the
shape sensing behavior of PAR proteins. Figure ~\ref{fig:ellipsoid}a shows the
top view of the ellipsoid as domains initialized close to $\theta =
0$ (Fig.~\ref{fig:ellipsoid}b) evolve over time. These domains move from the
ellipsoid center towards its narrow ends over a duration of 18 minutes. A domain
created just left of the center evolves to the left end, and a domain just right
of the center evolves to the right end. We formalize this by tracking the
angular position of the center of the high-$\rho$ domain
(Fig.~\ref{fig:ellipsoid}b). Varying the initial angular position, we find that
domains initialized closer to the left end, i.e. $\theta_\textrm{initial} \in
[0,\pi)$ end up at the left end ($\theta_\textrm{final} = \pi/2$), and all
domains initialized closer to the right end,  $\theta_{\textrm{initial}} \in
(\pi,2\pi]$ have $\theta_\textrm{final} = 3\pi/2$ (Fig. \ref{fig:ellipsoid}c).
This is precisely the binary shape sensing observed by \cite{mittasch_2018_non}.
We emphasize that our model is not a full model of the PAR system, which should
include the effect of antagonistic interactions between different PAR proteins
and hydrodynamic flow \cite{goehring_2011_polarization}; however, it illustrates
that minimalistic cell polarity models can capture binary shape sensing without
additional assumptions. 
The time taken to reach the steady state decreases from 1800 s for \(\theta \)
near 0 to 100 s for \(\theta\) near \(\pi/2\), and naturally increases
symmetrically as $\theta$ ranges from $\pi/2$ to $\pi$. This emphasizes the role
of the initial condition's symmetry: near-symmetric initial conditions can take
a long time for a spontaneous symmetry breaking to occur.

\begin{figure}[ht]
	\centering
	\includegraphics{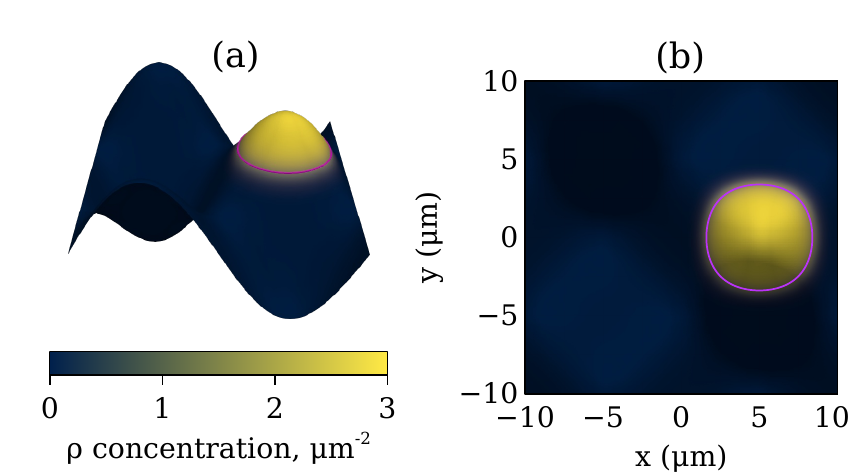}

	\caption{{\bf Steady state shape of a domain on a smooth sinusoidal test
	surface.} {\bf (a)} Steady-state solution of the reaction-diffusion
	equations on a sinusoidal surface. Magenta lines are shown to illustrate the
	boundaries of the high-$\rho$ domain marked at
	\(\rho=\left(\rho_\textrm{max}+\rho_\textrm{min}\right)/2\). {\bf (b)} Steady state of
	reaction-diffusion model (color map, viewed from above), is in agreement
	with the perimeter minimization (solid magenta line).}

	\label{fig:domaindef}
\end{figure}

\subsection{Wave-pinning and perimeter-minimization agree on a simple smooth
surface test problem}

\label{sec:smooth}

\begin{figure*}[bth]
	\centering
	\includegraphics{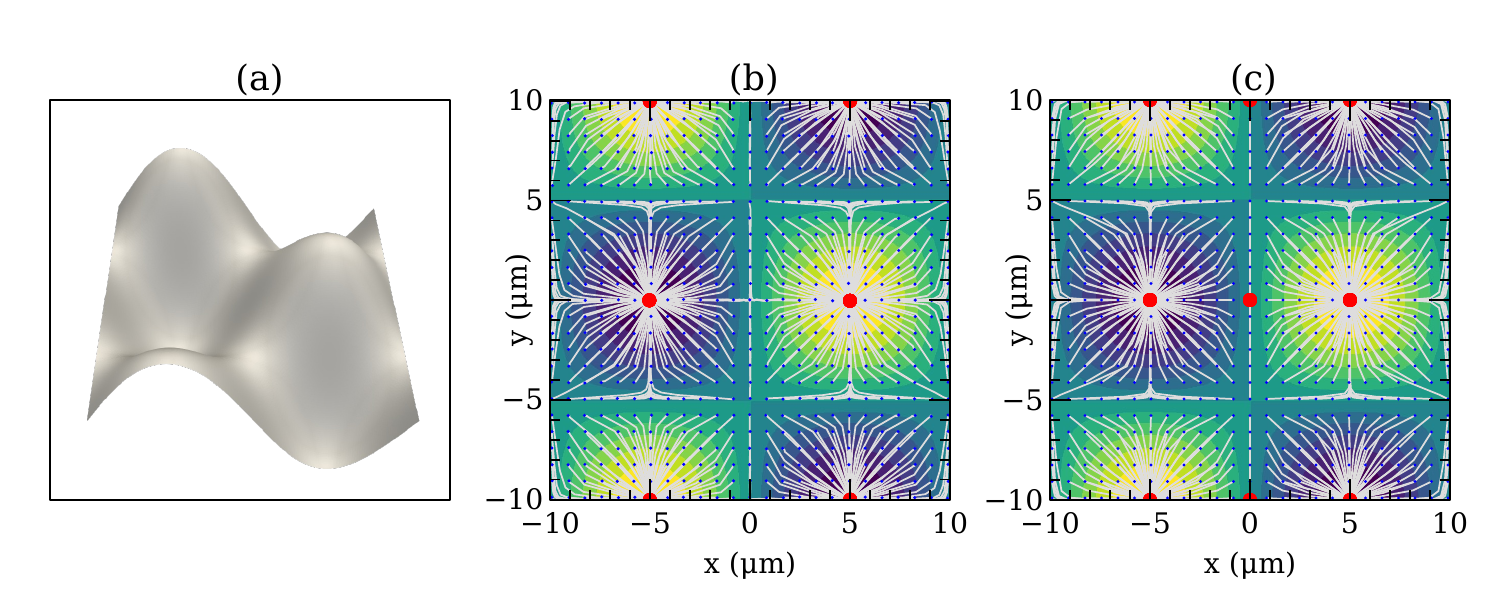}
	\caption{{\bf Domains on smooth surface robustly find peaks and valleys,
	with agreement between wave-pinning and perimeter minimization.} {\bf (a)}
	Rendering of the surface in \eq \ref{eq:smoothsurf}. {\bf (b)} and {\bf (c)}
	show the trajectories and steady-state positions of domains in the WP and PM
	models respectively. In both, {the uniformly spaced grid of small blue dots} indicate the initial position of a
	domain, the white line indicates the domain's trajectory, and heavy red dots
	are the domain's final location. Color maps are contours of the height
	$h(x,y)$.}

	\label{fig:r0steady}
\end{figure*}

{To understand how the dynamics of wave-pinning depend on
membrane shape in a simple context, we begin with a sinusoidal surface (\fig
\ref{fig:domaindef}a)
\begin{align}
 \label{eq:smoothsurf}
 h(x, y) = h_0\sin{\frac{2\pi x}{W_0}}\cos{\frac{2\pi y}{W_0}}
\end{align}
where $h_0 = 5.55 \,\um$ and $W_0 = 20 \,\um$, and the system spans $-W_0/2 \leq x,y \leq W_0/2$. The surface represents a portion of the cell membrane, and its
size and curvature are similar to that of the ellipsoid. {This surface has the
advantage of being able to be represented as a function $z = h(x,y)$, making
perimeter minimization simpler (see Appendix for numerical methods). }}

{We solve the reaction-diffusion equation \eq \ref{eq:rho} on
the simple sinusoidal surface. We find that, initializing the system with a
region of high \(\rho\) concentration near one of the surface peaks, the
high-activity domain migrates to the peak at long times (Fig.
\ref{fig:domaindef}a, {Movie 1}). {We simulate the reaction-diffusion process for 5,000 seconds to ensure we have converged to the steady state, but all but a few domains reach a steady state within a few hundred seconds (see, e.g. Movie 1).} The WP model gives as a result a concentration $\rho$ on
the surface that decreases rapidly but smoothly from a maximum value
$\rho_\textrm{max} \approx \rho_+$ to a minimum value $\rho_\textrm{min} \approx \rho_{-}$. To define an
explicit domain shape and size as in the perimeter minimization, we must choose
a threshold. We choose the contour of $\rho = \frac{1}{2}\left(
\rho_\textrm{min} + \rho_\textrm{max}\right)$ as the boundary for
calculating the area and shape of the high \(\rho\) concentration domain. 
\fig\ref{fig:domaindef}b shows the ``top'' view of the sinusoidal surface of
\fig\ref{fig:domaindef}a. The curve in magenta indicates the steady-state
shape of the domain obtained from perimeter minimization (\eq
\ref{eq:gradient_descent}), which agrees closely with the transition from high
to low $\rho$. This is consistent with our idea that perimeter minimization is a
good heuristic for explaining the shape-sensing behavior.}

\subsubsection*{Steady-state location of high activity domain depends on domain
initial position}

\begin{figure*}[htb]
	\centering
	\includegraphics{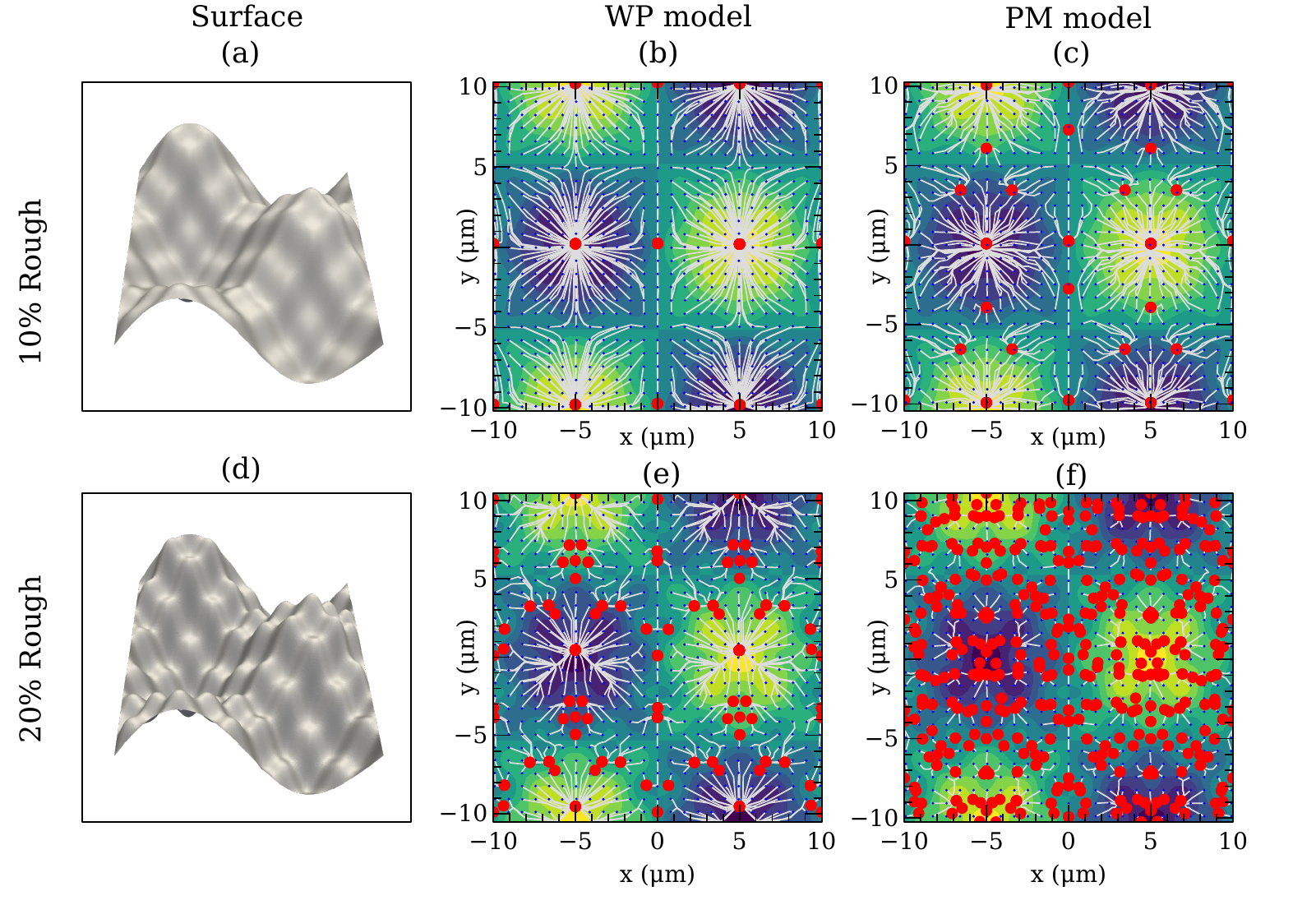}

	\caption{\textbf{Surface roughness disrupts shape-sensing as shown by the
	presence of steady states at locations away from the global peaks and the
	valleys of the surface. The wave-pinning model is more robust to it than
	perimeter minimization predicts as it has fewer such local steady states.}
	\textbf{(a)} Rendering of the surface in Eq.~\ref{eq:r_rough} with $h_1$ as
	10\% of $h_0$. \textbf{(b)} and \textbf{(c)} show the steady-state positions
	and the trajectories obtained from WP model and PM model respectively for
	the surface in Fig.~\ref{fig:roughSteadyState}a. \textbf{(d)} Rendering of
	the surface in Eq.~\ref{eq:r_rough} with $h_1$ as 20\% of $h_0$.
	\textbf{(e)} and \textbf{(f)} show the steady-state positions and the
	trajectories obtained from WP model and PM model respectively for the
	surface in Fig.~\ref{fig:roughSteadyState}d. As in Fig. \ref{fig:r0steady}, {the uniformly spaced grid of small blue dots} are domain initial positions, white lines are domain trajectories, and heavy red points are domain final positions. Colors indicate contours of $h(x,y)$. }

	\label{fig:roughSteadyState}
\end{figure*}

{Our results in Fig. \ref{fig:ellipsoid} and the experimental
results of \cite{mittasch_2018_non} show that regions of high active polarity
protein concentration on the membrane will evolve to different steady-state
locations depending on their initial location. What trajectory do they take? Our
simple test surface \(h(x, y)\) has multiple peaks and troughs as potential
steady-state locations for domains. 
To understand how domains choose their steady-state positions and what
trajectory they take, we simulate domains starting from 625 initial positions
shown as {a small grid of blue dots} in \fig\ref{fig:r0steady}. We do these simulations in both
the wave-pinning and perimeter minimization models.} The big red dots indicate
the steady state position of the center of mass of the domains. A thin white
line connecting a blue dot with a red dot indicates the trajectory of the center
of mass of the domain. {Generally, we observe that a domain
migrates to the peak or valley that is closest to it (Fig.~\ref{fig:r0steady}).
Interestingly, for points that are initialized nearly equidistant between two
peaks or valleys, these domains follow a trajectory tracing out the line of
symmetry between these points. The PM model and the WP model are in very good
agreement in both their predicted trajectories and steady-state domain locations. {While earlier work has suggested that WP models have a perimeter-minimizing property \cite{jilkine_2009_wave,maree2012cells,vanderlei_2011_computational}, it is somewhat surprising that the trajectories match this well between the PM and WP model, given the simplified overdamped dynamics we have chosen.}

There is a small disagreement between PM and WP for the three red dots along the central vertical line in the
minimization plot of \fig\ref{fig:r0steady}.} From the perspective of the
energy minimization, peaks and troughs of the sinusoid are equivalent.
Therefore, the central vertical line is a line of symmetry. Domains that start
with their center of mass exactly equidistant from a peak and a trough don't
migrate in the PM model, but do in the WP model, where the symmetry is broken at
a shorter timescale. The exact timescale of breaking a symmetry like this will
depend on both the details of the initialization and the rate of accumulation of
floating point errors in both models. Therefore, we would not necessarily expect
these domains initialized precisely on lines of symmetry to agree between WP and
PM. We argue that the apparent steady-states (red dots) away from the peaks and
valleys in \fig{\ref{fig:r0steady}} in the PM model are long-lived transients.
{Small perturbations away from these apparent steady states leads
to the domains migrating to the peaks and valleys, as with the other domains
(\fig\ref{fig:pm_transient}).}

\subsection{Shape sensing is disrupted on rough surfaces}

{Cell membranes have roughness due to the presence of filopodia,
blebs, embedded proteins, etc. Any mechanism for domain localization should be
robust to this roughness. Therefore, we would like to check if our models are
robust in predicting the steady states for rough surfaces. How rough can a
surface be before shape sensing by WP or PM breaks down? {We characterize the disruption of shape sensing by determining whether domains initialized to different locations on the membrane can still migrate to the global peaks and valleys when additional roughness is introduced.} To answer this question
we superimpose a small wavelength roughness of increasing amplitude on the
smooth sinusoidal surface, 
\begin{align}
 \label{eq:r_rough}
 h(x, y) = h_0\sin{\frac{2\pi x}{W_0}}\cos{\frac{2\pi y}{W_0}} 
    + h_1 \sin{\frac{2\pi x}{W_1}}\cos{\frac{2\pi y}{W_1}}
\end{align}
where $W_1 = 4 \, \um$ and we choose the amplitude of the perturbation $h_1$ to
be 10\% or 20\% of the amplitude $h_0$.}

\begin{figure*}[htb]
	\centering
	\includegraphics{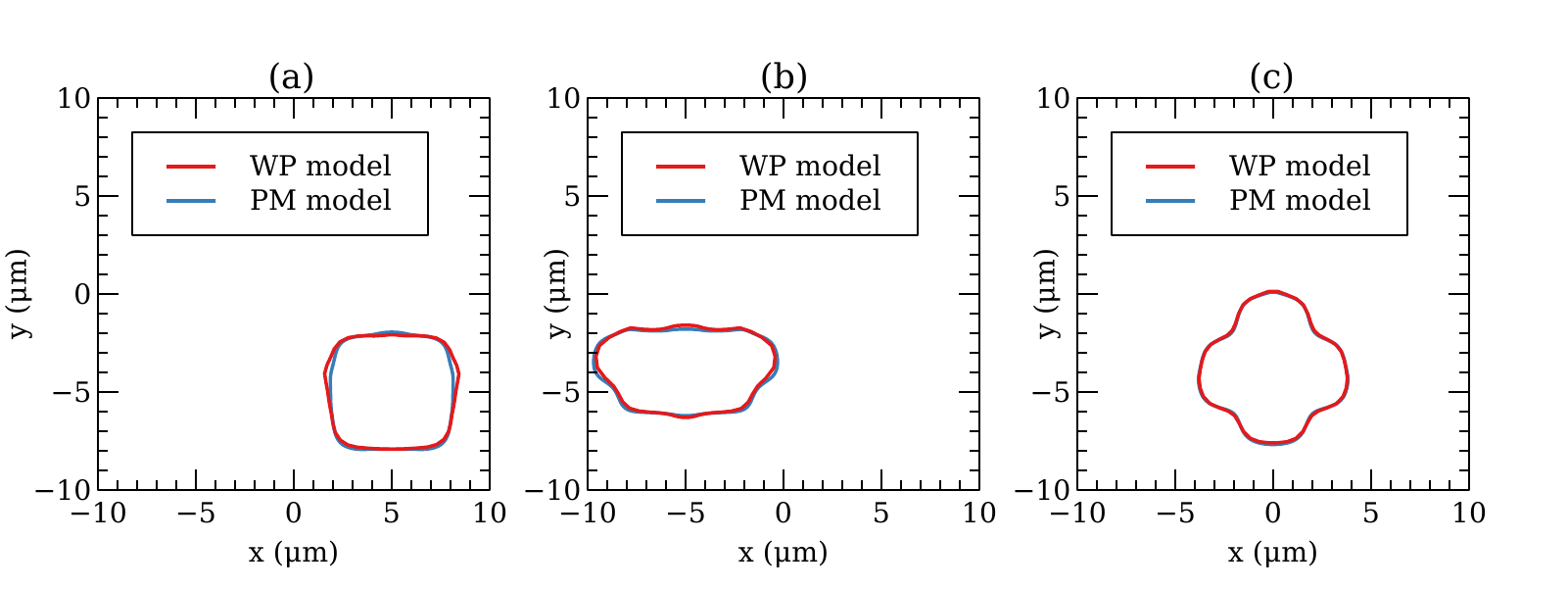}
	\caption{{\bf Steady-state domain shapes agree between wave-pinning and perimeter
	minimization even at 20\% roughness.} Three representative examples are shown from the subset
	of initial conditions where steady-state centroids are in agreement between WP and PM (see text).}
	\label{fig:wp_pm_shape_match}
\end{figure*}

We analyze the steady state positions of domains for the
same set of starting positions as studied in Fig.~\ref{fig:r0steady}. These results are somewhat involved, and are presented in Figs. \ref{fig:roughSteadyState}--\ref{fig:pm_area_fragile}.

\subsubsection*{Domain steady-states and trajectories}
\label{subsec:dsst}

{How are the domain steady states and trajectories altered in
the presence of roughness? \fig\ref{fig:roughSteadyState}b shows how domains
move in the WP model for a surface with 10\% roughness ($h_1 = 0.1 h_0$),
showing steady states as red dots. Most of the steady states are the same as in
\fig\ref{fig:r0steady} except for the three red dots at \(x=-5\
\mu\mathrm{m}\) and the three red dots at \(x=5\ \mu\mathrm{m}\), which lie along lines of symmetry
and can be long-lived transients as discussed above.  We find then that the WP model 
still localizes domains to the peaks
and valleys in the presence of a roughness of 10\% amplitude of the smooth
surface.}

{\fig\ref{fig:roughSteadyState}e shows the steady states of the WP model for
the 625 initial conditions for the surface with roughness 20\%. Now we see
several new steady states emerge in addition to the steady states of the smooth
surface. Again, we have some clusters of steady state around the lines of
symmetry at \(x=\pm 5\ \mu\mathrm{m}\) but there are other steady states which
are not near any lines of symmetry. Thus, the shape-sensing ability of the WP
model {-- in the sense of its ability to find the global peaks and valleys -- }has deteriorated for the increased roughness.}

{On a smooth surface, a domain that minimizes perimeter while
keeping area fixed localizes robustly to the peaks and valleys of the surface
and also reproduces the evolution of the WP model (\fig\ref{fig:r0steady}).
Both of these properties fail for a sufficiently rough surface.
\fig\ref{fig:roughSteadyState}c shows the solution of the PM model for the
10\% surface roughness. In comparison with \fig\ref{fig:r0steady} we note the
emergence of extra steady states, including states that are not near the lines
of symmetry and cannot be discounted as very
long-lived transients. We argue that the shape-sensing ability of the PM model
is affected more than the WP model for the same amount of surface roughness.}

{More dramatically, solving the PM model on a surface with 20\%
roughness shows a huge increase in the number of steady states
(Fig.~\ref{fig:roughSteadyState}f)
-- domains do not typically move over any significant distance on the surface
and are largely localized to near their initial position.  The {ability of the PM model to find global peaks and valleys} has broken down. As we noted above, in the WP model,
domains do not perfectly localize to the troughs and peaks at 20\% roughness,
but the level of new steady states created in PM at 20\% roughness is
qualitatively worse. This, again, suggests the relative robustness of the WP
model.}

\subsubsection*{Domain Shapes}

{For smooth surfaces, the steady states, trajectories, and shapes of domains are
identical between the perimeter minimization and wave-pinning models
(Fig.~\ref{fig:domaindef}b, \ref{fig:r0steady}). In the presence of surface
roughness, though, the steady states and trajectories do not match between WP
and PM (Fig. \ref{fig:roughSteadyState}). Does this indicate a complete failure
of the matching between the PM model and the WP model, or worse -- was the match
we saw in Fig. \ref{fig:r0steady} a coincidence? To check this question,
we test if the PM model and WP model can reproduce the same steady-state domain
shapes. However, it is only appropriate to compare the domain shapes at the same
location on the surface $h(x,y)$. For the 20\% roughness case, the number of
steady states predicted by the PM model (Fig.~\ref{fig:roughSteadyState}f) is
much larger than the number of steady states predicted by the WP model
(Fig.~\ref{fig:roughSteadyState}e), but for some of the 625 initial conditions,
the steady states of the WP model and the PM model have centroids that are close
to each other. We compare domain
shapes between PM and WP when the predicted steady-state domain centroids are
in agreement (within a tolerance of 0.15 \(\mu\mathrm{m}\)), finding that domain shapes match well between the models even in
the presence of 20\% surface roughness (Examples shown in
Figs.~\ref{fig:wp_pm_shape_match}a, \ref{fig:wp_pm_shape_match}b,
\ref{fig:wp_pm_shape_match}c). These shapes are nontrivial and complex,
very different from the circular domains found on the smooth surface, and the
agreement is excellent. Even though the PM model is unable to give the same
trajectories and the same steady states as the WP model, it is still quite
robust at reproducing the domain shapes.}

\begin{figure*}[htb]
	\centering
	\includegraphics{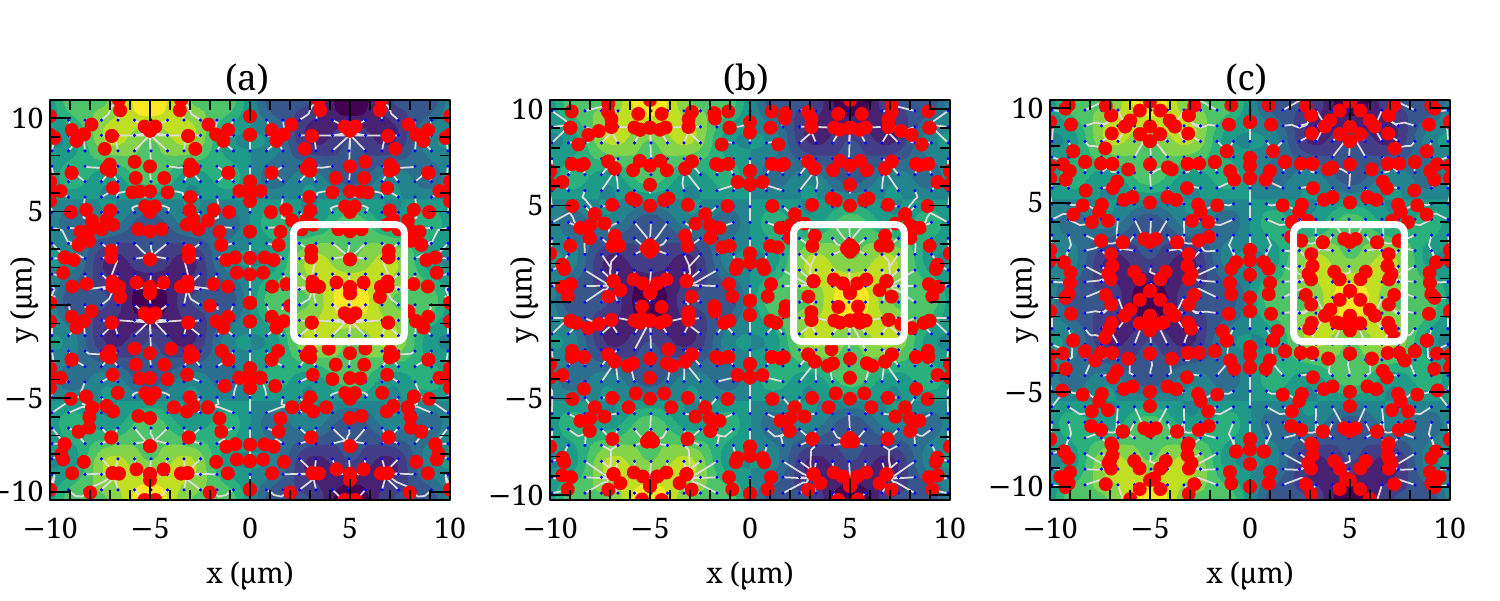}

  \caption{\textbf{Steady states predicted by perimeter minimization are
  sensitive to domain size.} {We have highlighted the steady states at the peak, marked with white squares, which is markedly different in the three subfigures.} \textbf{(a)},
  \textbf{(b)}, and \textbf{(c)} show the steady state positions and the
  trajectories given by the PM model when the domain area is set to 8\%, 9\%,
  and 10\% of the total surface area respectively. {As in Fig. \ref{fig:r0steady}, {the uniformly spaced grid of small blue dots} are domain initial positions, white lines are domain trajectories, and heavy red points are domain final positions. Colors are the contour map of $h(x,y)$.}}

  \label{fig:pm_area_fragile}
\end{figure*}

\subsection{Perimeter minimization on rough surfaces is fragile to small changes
in domain area}

{We saw in the previous section that perimeter minimization can
predict complex nontrivial domain shapes that appear in the WP model (Fig.
\ref{fig:wp_pm_shape_match}). This suggests that the steady states of WP do
obey a perimeter minimization principle -- at least locally. However, the
large-scale trajectories and many of the steady states differ between the PM
model and the WP model. Why? The PM model has a tight constraint on the domain
area, while in the WP model the domain area is only approximately fixed. In
fact, for the WP model the domain area varies slightly for different starting
positions of the center of mass and also at different locations of the center of
mass along the trajectory. For the simulations shown in \fig
\ref{fig:roughSteadyState}, we found that the domain area ranged from 8\%-10\%
of the total surface area of the rough membrane. Does perimeter minimization
predict the same domain steady state location or trajectory for these different
values of domain area?  Figures~\ref{fig:pm_area_fragile}a-c show the trajectory
plots obtained from the PM model when domain area is constrained to a value $A_0$
that is 8\%, 9\%, and 10\% of the total surface area for the surface with 20\%
roughness. The local minima predicted for different domain areas are different {(see, e.g. white boxes in Fig. \ref{fig:pm_area_fragile}a-c).}
This shows that strict perimeter minimization on a rough surface is fragile --
it depends so strongly on the target area $A_0$ that we should not expect
agreement between WP and PM.}

\begin{figure}[htb]
 \centering
 \includegraphics[width=\linewidth]{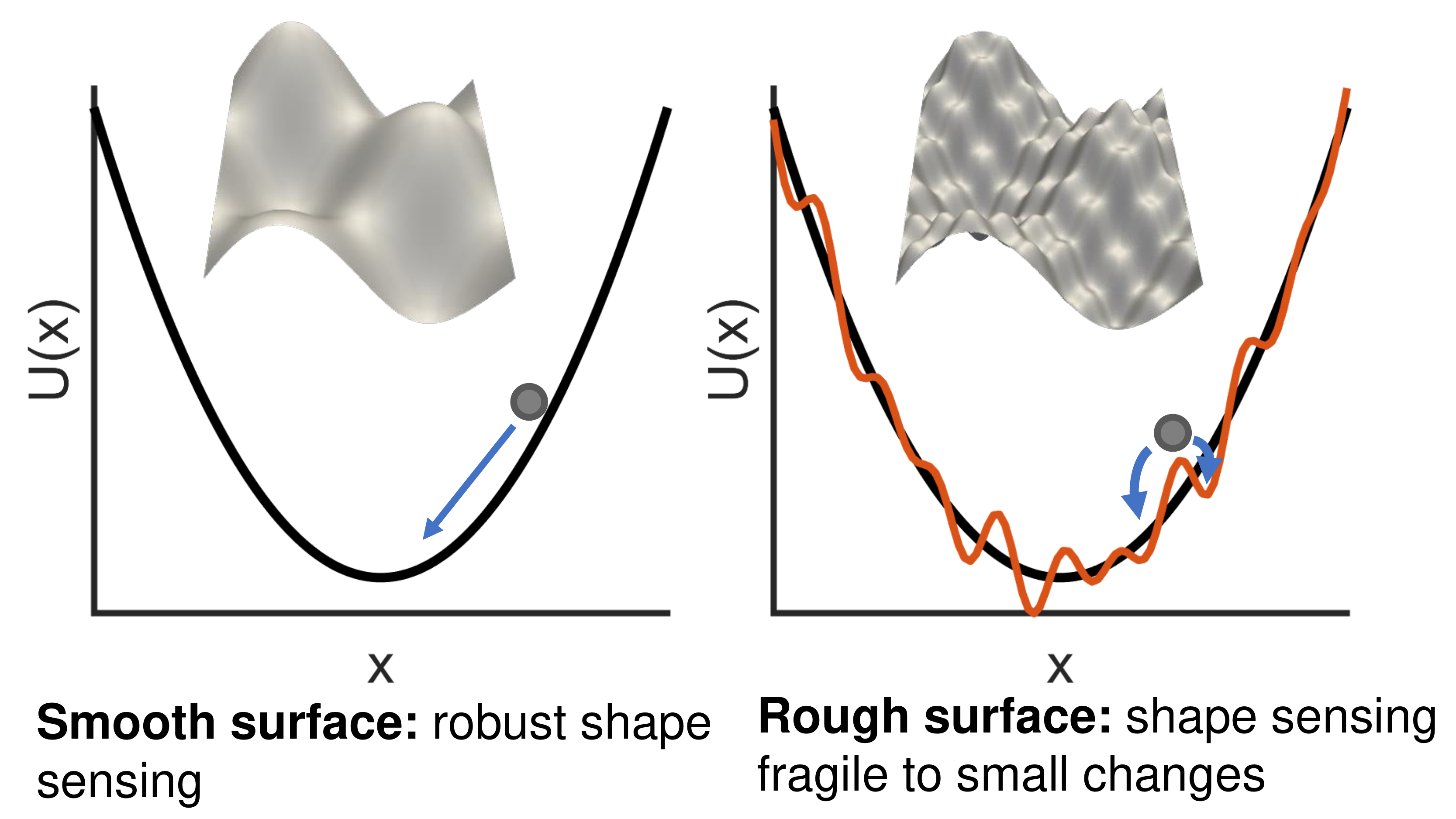}

 \caption{Movement of the domain over a surface is like moving a particle through energy landscape $U(x)$; shape-sensing is
 like finding the minimum energy position. For a smooth surface, the global
 minimum is easily attained but for rough surfaces there are local minima with
 energy barriers that prevent reaching the global minimum. Small changes to initial condition or the landscape parameters lead to changes in outcome (blue arrows, right)}

 \label{fig:schematic_roughness}
\end{figure}

{The fragility of perimeter minimization to small changes in
area can be understood by thinking about the domain as moving within an
effective energy landscape $U(x)$ (Fig. \ref{fig:schematic_roughness}). When the
surface $h(x,y)$ is smooth, this landscape is also smooth, and a domain can
smoothly travel to the peak or trough -- the energy minimum. However, when the
surface becomes rough, we expect the effective energy landscape to also become
rough (Fig. \ref{fig:schematic_roughness}, right) -- and domains become trapped
in local minima. More importantly for this section, we can see that small
perturbations to the landscape -- as might be expected from changing the domain
area -- can lead to large shifts in the steady state. This corresponds with the
fragility observed in Fig. \ref{fig:pm_area_fragile}.} 

\subsection{How reaction-diffusion shape sensing can be made robust to membrane
roughness}

The sensitivity of the PM model to the domain area does not fully explain the
difference between the steady states shown in \fig\ref{fig:roughSteadyState}
-- the WP model has many fewer unique steady-states than any of the PM results
presented in Fig.~\ref{fig:pm_area_fragile}. Why is the WP model
so robust?

\subsubsection*{Effect of diffusion coefficient}

The high \(\rho\) concentration domain formed by wave-pinning has a finite
interface width as shown in \fig\ref{fig:domaindef}a. The PM model domain, on
the other hand, has a sharp interface. The width of the interface increases with
the diffusion coefficient of the reaction-diffusion
system~\cite{hubatsch_2019_cell}. So as we decrease the diffusion coefficient,
the interface width of the WP model will decrease -- potentially leading to better agreement between WP and PM. 

\begin{figure*}[htb]
	\centering
	\includegraphics{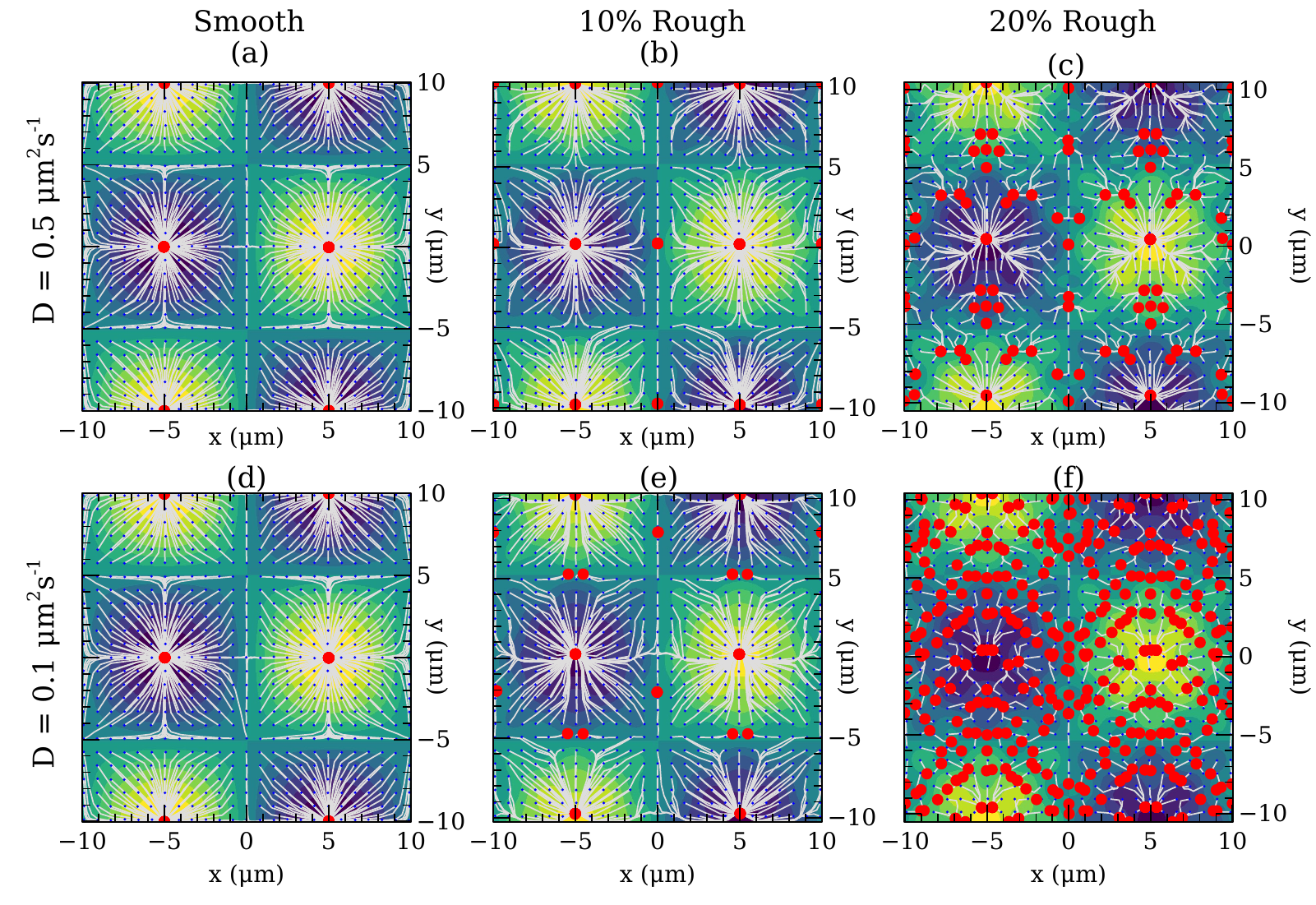}

	\caption{\textbf{In the WP model, the width of the interface of high
	activity domain can be decreased by reducing the diffusivity. As the WP
	model approaches the sharp interface limit, many local minima steady states
	appear.} {\bf (a)}, {\bf (b)}, and {\bf (c)} show the steady state positions
	for \(D = 0.5\mathrm{\mu m^2s^{-1}}\) for increasing surface roughness. {\bf
	(d)}, {\bf (e)}, and {\bf (f)} show the corresponding steady state positions
	for \(D = 0.1\mathrm{\mu m^2s^{-1}}\) for increasing surface roughness. As in Fig. \ref{fig:r0steady}, {the uniformly spaced grid of small blue dots} are domain initial positions, white lines are domain trajectories, and heavy red points are domain final positions. Colors are contour map of $h(x,y)$.}

	\label{fig:diffusion_coefficient_vary}
\end{figure*}

For the smooth surface, the WP model and the PM model are in good agreement (see
\fig\ref{fig:r0steady}), and reducing the diffusion coefficient in the WP
model does not alter the results for this surface
(Fig.~\ref{fig:diffusion_coefficient_vary}a,d).

In our default parameters, 
domains in the WP model largely reach the peaks or valleys of the surface at 10\% roughness (Fig. \ref{fig:roughSteadyState}b). However, on reducing the
diffusion coefficient from \(0.5\ \mu\mathrm{m\,s}^{-1}\) to \(0.1\
\mu\mathrm{m\,s}^{-1}\), the WP model begins to show nontrivial local minima (Fig. \ref{fig:diffusion_coefficient_vary}e), albeit
different from those shown by the PM model as shown in
\fig\ref{fig:roughSteadyState}.
As we increase the surface roughness to 20\% for $D = 0.1\
\mu\mathrm{m\,s}^{-1}$, the number of local minima for the smaller
diffusion coefficient increases significantly (Fig.~\ref{fig:diffusion_coefficient_vary}f). The number of local minima for \( D = 0.1\
\mu\mathrm{m\,s}^{-1}\) for the 20\% roughness case are comparable to the number
of local minima seen in the PM model (see \fig\ref{fig:roughSteadyState}f).

{We emphasize that our results in Fig.
\ref{fig:diffusion_coefficient_vary} show that decreasing the diffusion
coefficient increases the number of steady states -- this is not just an
artifact of decreasing diffusion making kinetics slower. Because of the slower
diffusion coefficient, we ran the simulations in \fig
\ref{fig:diffusion_coefficient_vary} with \(D = 0.1 \mu m^2/s\) for a time of
\(20000\) s, compared with \(5000\) s for \(D = 0.5 \mu m^2/s\). We also ensured
that these steady states were converged by} {running the
simulations for finer finite element meshes, and for longer simulation times for
a subset of the initial positions.}

We argue that the robustness of the WP model as compared to the PM model is due
to the finite interface width of the high \(\rho\) concentration domain, and
that we can control whether the WP model seeks the true peaks and valleys or
gets stuck in local minima in part by changing this interface width via $D$.

\subsubsection*{Effect of domain size}

Along with size of the roughness, and the size of the interface, the diameter of
the high \(\rho\) concentration domain is another length scale in this problem.
{Can cells sense shape more effectively when the domain is
probing a larger length scale? We investigate the effect of size of the high
\(\rho\) concentration domains on the steady states by increasing \(M/S\), which
{strongly influences domain size}, from \(2.9\ \mu \mathrm{m}^{-2}\) to \(3.2\ \mu
\mathrm{m}^{-2}\). This increases the typical domain size to \(\sim\)18\% of the
surface.}

\begin{figure}[htb]
	\centering
	\includegraphics{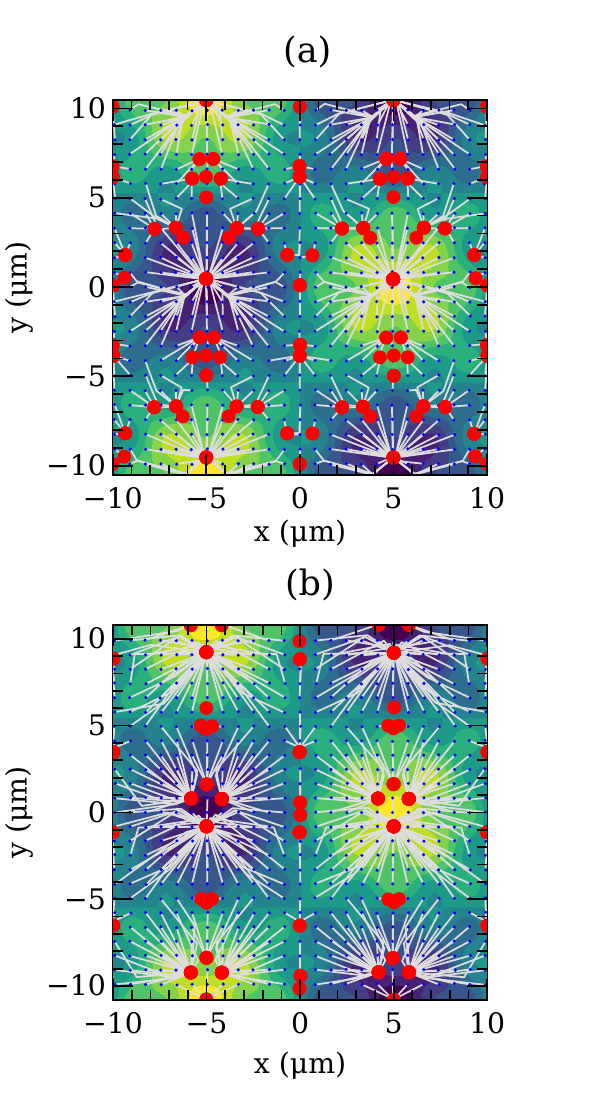}

	\caption{\textbf{Increasing the size of high activity domains enhances the
	shape-sensing ability of the WP model. The steady state positions in (b) are
	closer to the global peaks and valleys.} {\bf (a)} shows the steady states
	and trajectories obtained from the WP model when $M/S = 2.9 \mu m^{-2}$ and the size of the domains is
	9\% of the total surface area. {\bf (b)} shows the steady states and
	trajectories obtained from the WP model when $M/S = 3.2 \mu m^{-2}$ and the size of the domains is 18\%
	of the total surface area. As in Fig. \ref{fig:r0steady}, {the uniformly spaced grid of small blue dots} are domain initial positions, white lines are domain trajectories, and heavy red points are domain final positions. Colors are a contour map of $h(x,y)$.}
	\label{fig:d5r20a18steady}
\end{figure}
For the smooth surface and the surface with 10\% roughness, there were no local
minima in the WP model (see \fig\ref{fig:r0steady}) and
\fig\ref{fig:roughSteadyState}b). Increasing the domain size did not produce
any significant difference in these two cases. But for the surface with 20\%
roughness, some of the local minima that were not situated on any line of
symmetry appear to ``coalesce'' as shown in \fig\ref{fig:d5r20a18steady}. At
the lines of symmetry, as we have already noted, it is difficult for the domains
to localize to a unique solution. So, in general, it seems that increasing the
domain size, increases the robustness of the shape-sensing slightly.

\subsection{{Shape sensing occurs in a broad range of biologically-relevant geometries}}
\begin{figure*}[htb]
	\centering
	\includegraphics[width=\textwidth]{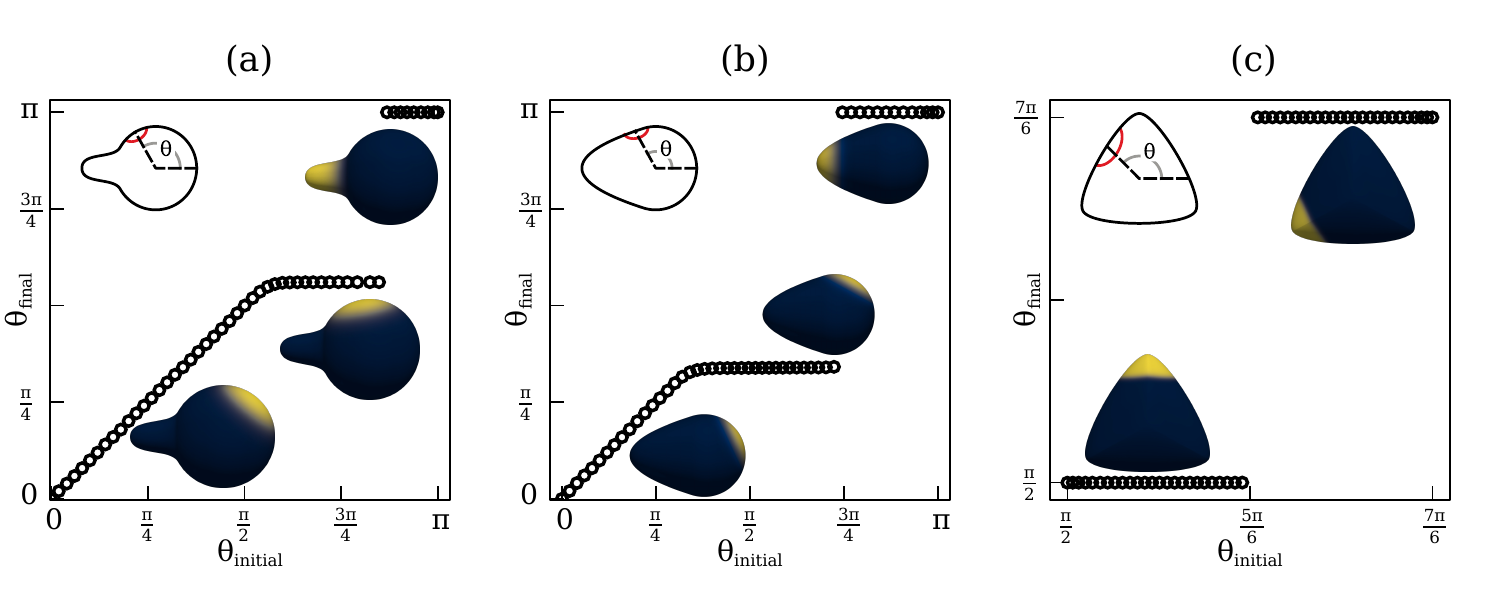}

	\caption{{\textbf{Shape sensing from wave-pinning occurs in a broad range of biologically-relevant three-dimensional cell shapes.}  We show the final angular position of a domain as a function of its initial angle, as in Fig. \ref{fig:ellipsoid}c. $\rho$ concentration is plotted for representative examples. Cell shapes are a) spheres with a bump (``shmoo"), b) teardrop, and c) a samosa shape modeling a cell in a triangular confinement \cite{klinkert_2019_aurora}. Domains initialized at one location on a three-dimensional cell surface systematically evolve to their nearest steady-state location.}}
	\label{fig:newshapes}
\end{figure*}

{The insight we have developed in the simple sinusoidal surface can help us understand both past computational work and experiments beyond our motivating example of binary shape sensing \cite{mittasch_2018_non}. We show the evolution of the wave-pinning reaction-diffusion model on three different 3D surfaces in Fig. \ref{fig:newshapes}: a sphere with a bump and a sphere with a larger bump, both approximating the shmoo-like structures of mating yeast as previously simulated in, e.g. \cite{orlandini_2013_domain,trogdon2018effect}, and a cell with triangular symmetry. This last shape models C. elegans zygotes studied in triangular confinement by \cite{klinkert_2019_aurora}. We have argued above that, on smooth surfaces at least, the high-activity domain evolves to locally minimize the perimeter while keeping a fixed area. What would perimeter minimization predict on these surfaces? Let us think about a perfect sphere first. Given the rotational symmetry of a sphere,  no angle is preferred, so a domain initialized in one location will stay in that location, as if it were on a flat surface. Similarly, because the sphere with a bump is locally identical to a sphere, except in the immediate vicinity of the bump, we expect domains to remain at their initial centers of mass. We then simulate the wave-pinning equations for a sphere with a bump in Fig. \ref{fig:newshapes}a. When we initialize domains at different angles with respect to the bump, we find that domains on the larger spherical region remain at their initial angle. However, when domains are initialized closer to the bump, they are repelled by the bump -- localizing to the nearest undistorted portion of the sphere. This is similar to the final localization observed in \cite{orlandini_2013_domain} in a similar geometry (Fig. 7A in that paper). However, once the domain is initialized sufficiently close to the bump, it localizes to a final position on the bump -- the local minimum of perimeter with fixed area. This ``sphere with a bump'' geometry is particularly informative because it has a large region that is identical to an unperturbed sphere -- leading to a large fraction of the surface where domains will not migrate significantly. This shows the value of understanding the dynamics in terms of a {\it local} minimization of perimeter: even though there is a ``global'' location with smaller perimeter -- placing the domain at the bump -- domains initialized on the sphere do not migrate. This may explain why shape sensing in related reaction-diffusion models was viewed as weak \cite{orlandini_2013_domain}. By contrast, if the region of the bump is increased (Fig. \ref{fig:newshapes}b), domains become attracted to the peak over a larger range of initial conditions. These results show how shape sensing can be disguised on surfaces that are sufficiently close to a sphere. Our results also illustrate that it is essential to study a range of different initial conditions: there are often many different steady states even on a simple surface, and a full understanding of shape sensing requires seeing under which conditions different initial conditions converge to these steady states.}

{More excitingly, we see that the same mechanism of shape sensing by wave-pinning that we have studied above can explain additional elements of domain localization. In Fig. \ref{fig:newshapes}c, we simulate the wave-pinning reaction-diffusion model on a cell with triangular symmetry, similar to the shapes of zygotes confined in triangular wells studied by \cite{klinkert_2019_aurora}. In that paper, the authors found that in zygotes depleted of AIR-1, PAR-2 localized to the triangular corners of the cell -- the regions of high curvature -- and were able to reproduce this with a model in which the rate of binding to the membrane was sensitive to curvature. Here, we show that curvature-dependent binding is not necessary to reproduce localization to the corners. If we initialize a domain to the sides of the cell, we observe that it migrates to the corner (Fig. \ref{fig:newshapes}c). This is, again, consistent with our intuition from the perimeter minimization idea: domains with the same area will have a lower perimeter if they are localized to the tips. (We note: for the corners to have a lower perimeter, there must be some Gaussian curvature at the corners of the cell. We would not expect corner localization if the cell were shaped, for instance, like a triangular prism.) Corner localization behavior would not have been seen in the original simulations of \cite{klinkert_2019_aurora} in the absence of curvature-dependent binding rates $k_\textrm{on}$, because they worked in one dimension -- which does not resolve the full shape of the domain's perimeter. While the wave-pinning model does not serve as a complete model for the PAR system, it shows that a minimalistic cell polarity mechanism can reproduce corner localization without any additional assumptions.}

\section{Discussion}

{In this paper, we have shown that the classic wave-pinning
model of cell polarity is sufficient to produce the behavior of binary shape
sensing observed by \cite{mittasch_2018_non} -- as might be anticipated by
earlier work
\cite{camley_2017_crawling,cusseddu_2019_coupled,vanderlei_2011_computational}.
We then try to capture some of the essential features and limitations of shape
sensing by the wave-pinning Rho GTPase dynamics as well as developing a
heuristic model for it in terms of minimization of an effective energy
proportional to the domain perimeter. Wave-pinning reliably senses cell shapes
when the cell is smooth, but introduction of surface roughness significantly
disrupts this process. The ability of wave-pinning to sense cell shape in the
presence of roughness is controlled by both the diffusion coefficient $D$ on the
surface (\fig \ref{fig:diffusion_coefficient_vary}) and the domain size (\fig
\ref{fig:d5r20a18steady}). The perimeter minimization heuristic captures much of
the dynamics of domain migration, but fails on rougher surfaces (\fig
\ref{fig:roughSteadyState}) due to the relevance of finite interface sizes and
domain area fluctuations (\fig \ref{fig:pm_area_fragile}).} {However, the perimeter minimization heuristic {\it does} let us understand when domains on more complex, biological shapes should migrate, including predicting the corner-localization of PAR domains in triangular confinement (Fig. \ref{fig:newshapes}.}

How fast is shape sensing by wave pinning? Does it occur on a
biologically relevant timescale? In the experiments on long-axis polarization of
PAR proteins\cite{mittasch_2018_non}, the high-PAR concentration domain reaches
a steady states in approximately 10 minutes. Our modeling shows that protein
domains attain their steady states on a timescale of 100-1800 seconds,
compatible with a roughly 10 minute timescale. The longest times required to
reach steady state occur when the domain is initialized nearly symmetrically,
requiring a symmetry breaking. The timescale is similar between the
three-dimensional ellipsoidal domain and our smooth sinusoidal surfaces, where
except for those initial conditions at lines of symmetry, domains take $\sim$
100-500 s to reach steady state. Earlier work from Cusseddu \textit{et al.} have
reported results for WP simulations on a capsule~\cite{cusseddu_2019_coupled},
similar to the ellipsoidal shape of
Section~\ref{sec:results}\ref{subsec:bistability}, finding a much longer time of
13463 s to reach steady state. This may have been influenced by a symmetric
initial condition. However, unlike their simulation model, we have assumed that
the interior is well-mixed, and we have also chosen a larger diffusion
coefficient of 0.5 \(\mu\)m\(^2\)s\(^{-1}\) compared to their value of 0.1
\(\mu\)m\(^2\)s\(^{-1}\). Koo \textit{et al}~\cite{Koo_2015}
have reported \textit{in vivo} experimental results that Rho GTPases show six
different diffusive states of which the average diffusivity of the most
probable diffusive states is closer to \(0.5\)\(\mu\)m\(^2\)s\(^{-1}\), though we note that our model does not yet address the possibility of multiple states with different diffusion coefficients. The diffusion coefficients and kinetics may influence the plausibility of shape sensing in different contexts. If the time to attain the steady state is as long as 13000 seconds,
the large-scale shape of the cell will likely change due to other phenomena, like formation of
protrusions, before shape-sensing by wave-pinning happens. {Earlier work finding relatively weak effects of membrane shape on polarization \cite{orlandini_2013_domain} used membrane diffusion coefficients based on those for yeast, $D \approx 0.0025 \mu m^2/s$ \cite{goryachev2008dynamics}; this value is orders of magnitude smaller than in our case, and may lead to very slow, if any, shape sensing. Future work incorporating reaction-diffusion on a moving membrane (following, e.g. \cite{tamemoto_2020_pattern,tamemoto_2021_reaction}) would be required to understand whether shape sensing that is slow relative to surface motion would probe the time-averaged surface or be disrupted by dynamic changes.}

{Our results show that, at least for smooth surfaces, the
reaction-diffusion WP model is well-captured by perimeter minimization. This
result is consistent with earlier work relating mass-conserved reaction
diffusion equations and coarsening driven by interfacial tension (i.e. perimeter
minimization) in simpler contexts
\cite{tateno2021interfacial,bergmann2018active,brauns2020phase}, as well as the
analysis of \cite{jilkine_2009_wave}.  However, our work highlights crucial
limitations of these assumptions. The fragility of energy minimization on rough
surfaces to small changes in area (\fig \ref{fig:pm_area_fragile}) shows that an
approximate conservation of domain area is not sufficient to completely
characterize domain trajectories and steady states, and that the full
reaction-diffusion equations need to be solved. In addition, we note that the WP
model consists of two stable phases only when the concentration of the inactive
form is in a suitable range~\cite{mori_2008_wave,cusseddu_2019_coupled}. The
ability of cells to sense shape, as with their ability to polarize
\cite{hubatsch_2019_cell}, will be dependent on cell size and total Rho GTPase
amount.}

{Throughout this paper, we have assumed that the cytosolic component of the Rho GTPase is well-mixed and therefore uniform. This contrasts with
the key role of the cytosolic diffusion proposed in long axis selection by
\cite{gessele2020geometric}, and the full bulk-surface implementation of the wave-pinning model in \cite{cusseddu_2019_coupled}. The well-mixed cytosol is a common assumption, given the large difference in cytosolic and membrane-bound diffusion coefficients, but it is a  potential limitation of the model. If we extended our model to allow for a finite
level of cytosolic diffusion, the cytosolic inactive form might not be homogeneous, as observed previously \cite{maree2012cells}. With an inhomogeneous cytosol,
the local ratio of surface area to volume might
play a significant role \cite{meyers_2006_potential}. This would break the
symmetry in our model between positive and negative curvature -- peaks and
valleys would no longer be identical from the point of view of the model.}

{Our approach shows that shape sensing may emerge from wave
pinning without any explicit dependence upon membrane curvature in protein
binding or kinetics. Previous work has suggested that in order to explain PAR domain localization in C. elegans zygotes in triangular confinement, binding rates in the reaction-diffusion model must be dependent on curvature \cite{klinkert_2019_aurora}. We have shown in Fig. \ref{fig:newshapes} that the corner localization of these domains does not require this curvature-dependent binding, but can be reproduced solely from the minimal reaction-diffusion wave pinning model if simulated in a three-dimensional geometry. If we extended our model to study explicit dependence of binding rates
on local curvature, this dependence could also be used to break the symmetry between positive and negative curvature, as with cytosolic diffusion. However, the ability of single
proteins to sense micron-scale curvature on their own is rare, and many aspects
of the mechanism of micron-scale curvature sensing remain unresolved
\cite{cannon_2017_unsolved,cannon_2019_amphipathic}. We are not aware of any
evidence showing that Rho GTPases or PAR proteins have binding rates that depend on local
curvature. We therefore suggest that preferences for different signs of
curvature are more likely to arise from cytosolic diffusion effects.}

{Our results show that surface roughness can impede shape
sensing by both energy minimization and Rho GTPase dynamics, with perimeter
minimization more strongly affected. How crucial is the effect of roughness to
understanding shape sensing in realistic geometries? This likely varies between
cell types: the C. elegans zygote appears fairly smooth on the micron scale
\cite{mittasch_2018_non},  while at the other extreme, blebby cell surfaces may
have overhangs and extremely complex involutions \cite{driscoll_2019_robust},
for which the mechanism of {domain migration shape sensing studied here} seems implausible. The
importance of roughness is also dependent on the diffusivity of the
membrane-bound Rho GTPases (\fig \ref{fig:diffusion_coefficient_vary}).
Diffusion of polarity proteins could be altered by their coupling to the
cytoskeleton or other proteins
\cite{gowrishankar2012active,kusumi2011hierarchical,hosaka2017lateral,swartz2021active},
controlling the extent to which shape sensing leads to domains finding the
global peaks and valleys or long axis, or rather becoming pinned to a local
minimum. It is also possible to regulate the size of polarity domains by
changing the available concentration of Rho GTPases; this will also alter shape
sensing (\fig \ref{fig:d5r20a18steady}).} 

Our results clarify both the power and limits of shape sensing by reaction-diffusion and perimeter minimization: in the best case, shape sensing proceeds in a straightforward, predictable way, and robustly finds the minima and maxima in reasonable amounts of time. However, both mechanisms are fragile to sufficiently rough perturbations. This suggests that many past models of spontaneous cell turning, cell polarization, etc., \cite{camley_2017_crawling,goehring2011polarization,edelstein2013simple,mogilner2012cell,holmes2017mathematical} may need to be systematically tested to determine to what extent they are robust to realistic changes in cell geometry. 

\section*{Code availability}
Simulation code for reaction-diffusion and perimeter minimization is posted on Zenodo: \href{https://zenodo.org/record/6731244}{https://zenodo.org/record/6731244}

\begin{acknowledgments}
BAC acknowledges support from the NSF grant DMR-1945141. This research project was conducted using computational resources at the Maryland Advanced Research Computing Center (MARCC).
\end{acknowledgments}

%

\onecolumngrid
\appendix

\section*{{\Large SI Appendix}}

\renewcommand\thefigure{S\arabic{figure}}    
\setcounter{figure}{0}
\renewcommand\thetable{S\arabic{table}}    
\setcounter{table}{0}

\section{Movie captions}

Movie 1: Evolution of wave-pinning model on the sinusoidal surface showing the migration of a domain from its initial position to the final position at the surface peak. View is from the top.

\section{Parameter Values}

We tabulate the default parameter values used in our simulations. Any changes
from the default values have been explicitly mentioned in the main text.
\begin{table}[htb]
	\centering
	\begin{tabular}{|c|l|l|}
		\hline
		Parameter & Default Value & Description \\
		\hline
		\(k_0\) & 0.07 \(\mathrm{s}^{-1}\) & Basal activation rate \\
		\(\gamma\) & 5 \(\mathrm{s}^{-1}\) & Positive feedback activation rate \\
		\(K\) & 2  \(\mu\mathrm{m}^{-2}\) & Saturation parameter \\
		\(\delta\) & 3 \(\mathrm{s}^{-1}\) & Deactivation rate \\
		\(D\) & 0.5 \(\mu\mathrm{m}^{-2}\mathrm{s}^{-1}\) & Diffusion coefficient of the active \(\rho\)\\
		\(M/S\) & 2.9  \(\mu\mathrm{m}^{-2}\) & Total concentration (see Eq.~\ref{eq:rhocyt})\\
		\(S_{R0}\) & 627 \(\mu\mathrm{m}^2\) & Total surface area of the smooth surface\\
		\(S_{R10}\) & 664 \(\mu\mathrm{m}^2\) & Total surface area of the surface with 10\% roughness\\
		\(S_{R20}\) & 768 \(\mu\mathrm{m}^2\) & Total surface area of the surface with 20\% roughness\\
		\(A_{R0}\) & 0.09 \(S_{R0}\) & Surface area constraint for PM model on the smooth surface\\
		\(A_{R10}\) & 0.09 \(S_{R10}\) & Surface area constraint for PM model on the 10\% roughness surface\\
		\(A_{R20}\) & 0.09 \(S_{R20}\) & Surface area constraint for PM model on the 20\% roughness surface\\
		\(k\) & \(10^3\ \mu\mathrm{m}^{-1}\) & Penalty coefficient (see Eq.~\ref{eq:perimin})\\
		\(\beta\) & \(10^{-2}\ \mu\mathrm{m}\) & Gradient-descent step size coefficient (see Eq.~\ref{eq:gradient_descent})\\
{$T_\textrm{sim}$} & 5,000 s & Simulation time ($D = 0.5 \mu m^2/s$ simulations) \\
        {$N_{\textrm{elem}}$} & 80000 & {Number of mesh elements} \\
		\hline
	\end{tabular}
	\caption{Default values of the parameters used in the simulations of the PM and the WP models.}
	\label{table:param}
\end{table}

\section{Transient States in Perimeter Minimization}

\begin{figure}[h!]
	\centering
	\includegraphics{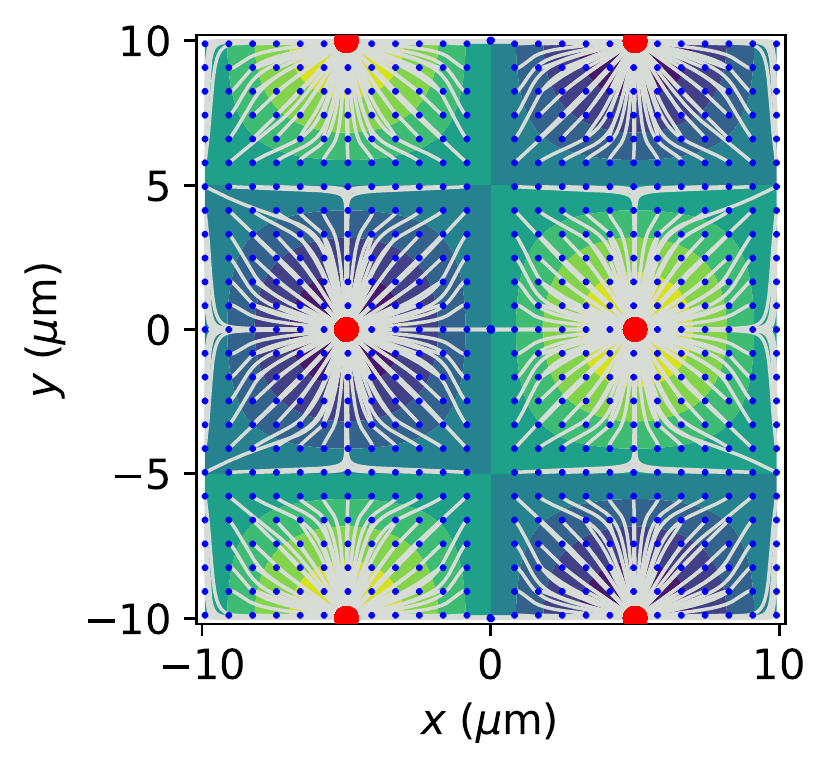}
	\caption{The steady states and trajectories obtained from the perimeter
	minimization simulations on the smooth surface after applying a small
	perturbation to the final states along \(x=0\) line shown in
	\fig\ref{fig:r0steady}.}
	\label{fig:pm_transient}
\end{figure}

In \fig\ref{fig:r0steady}c, we noted that perimeter minimization predicts some
three ``steady'' state along the \(x=0\) line that have been marked as red dots
in the plot. The initial positions of the centroid of the domains in all of
these cases were along the \(x=0\) line which is a line of symmetry. If we apply
a small random perturbation of 0.02 \(\mu\)m to each point \((x_i, y_i)\) of the final states of these
simulations, they attain steady states on one of the peak or valleys. Thus, we
confirm that these are indeed transient states and are an artefact of the
symmetry of the surface. \fig\ref{fig:pm_transient} show the trajectory plot
after perturbing the final states.

\section{Details of the Numerical Methods}

\subsection{Finite Element Model for Wave-pinning on a 3D surface}

Our approach uses some standard finite element tools: this section will be easier to follow with knowledge of weak forms of
PDEs, discretization of the domain of PDEs into a linear triangular mesh, the
shape functions of a linear triangular element, and Gaussian quadrature on
linear triangles to evaluate integrals defined over the elements.

\begin{figure}[htbp]
	\centering
	\includegraphics[width=0.5\textwidth]{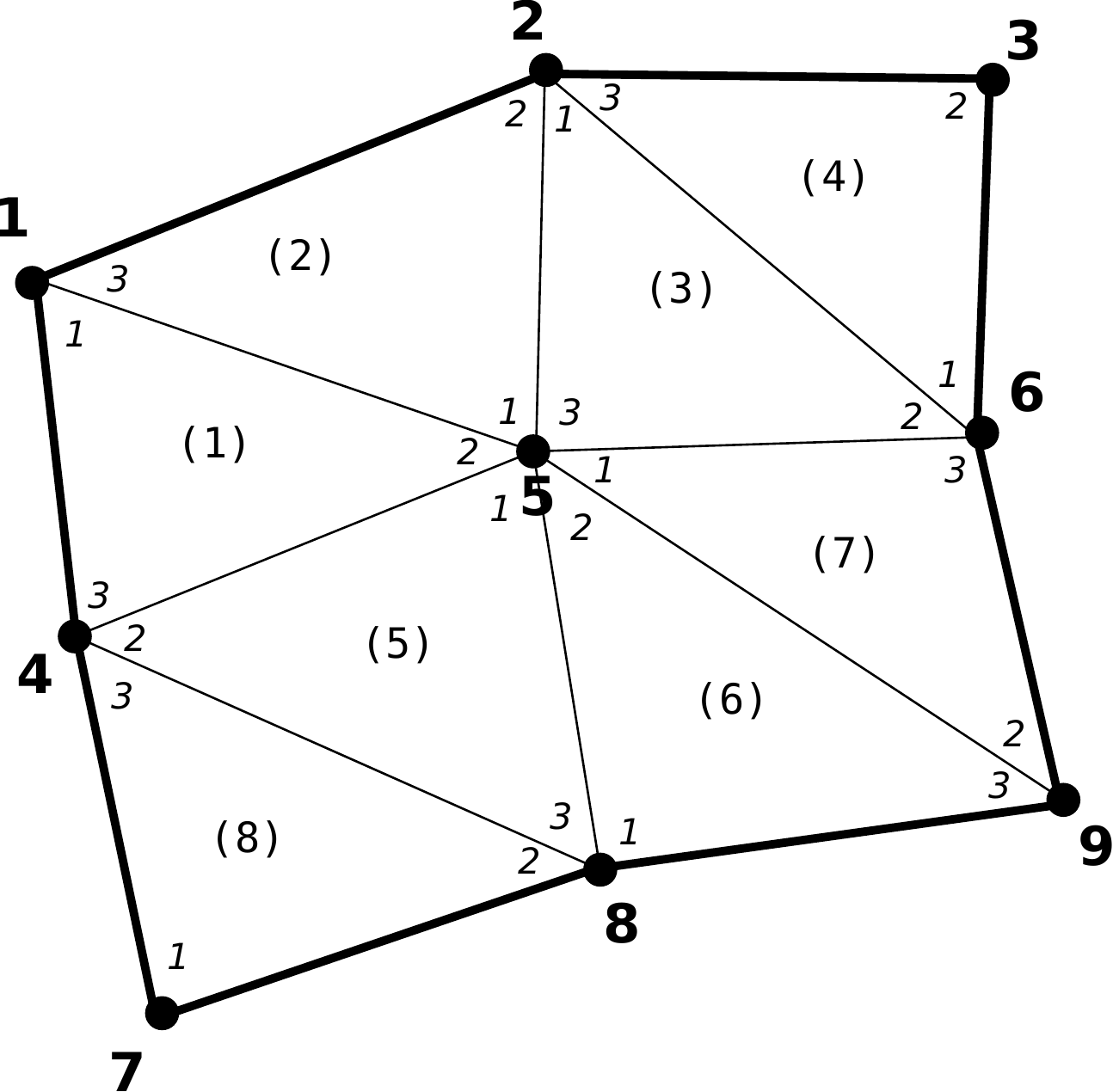}
	\caption{A sample surface \(\Omega\) which is the domain of a PDE to be
	solved using FEM is shown with its boundaries as bold lines. It has been
	discretized into a mesh of non-overlapping linear triangles \(\Omega_e\)
	where \(e=1,2,\ldots,8\) shown with thin lines. The circular dots are the
	nodes of the mesh that are coincident with the vertices of the triangles.
	The bold numbers show the global indices of the nodes. The italic numbers
	show the local indices of the nodes in a triangle. The number in parenthesis
	indexes the triangles.}
	\label{fig:mesh}
\end{figure}

We want to solve the following system of equations
\begin{align*}
\pdv{a}{t} &= D\laplacian a + f(a, b)\\
b &= \alpha - \frac{1}{S}\int_{\Omega} a
\end{align*}
using the Finite Element Method. Here, for conciseness, we've written the
membrane-bound concentration as $a$ and the uniform cytosolic concentration as
$b$. $\alpha = \frac{M}{\omega S}$. We've also implicitly assumed $\omega = 1
\mu m$ here, but this can be generalized to nonzero $\omega$ by taking
$S\to\omega S$ in the final equations. The first step for setting up the
finite-element problem is to derive the ``weak form'' of the PDEs.  We can
express the time derivative using finite-difference. Let $a$ be the
concentration at time-step $n$ and $\bar{a}$, be the value of $a$ at time step
$n -1$. Let the time step size be $t$.
\begin{align*}
a - \bar{a} = tD\laplacian a + tf(a, b)
\end{align*}
We will linearize the non-linear reaction function $f(a, b)$ using $a^{k+1} =
a^{k} + \delta a$ and $b^{k+1} = b^k + \delta b$. Here $\delta a$ is the
correction to the value of $a^{k}$ at the $k$th Newton iteration and
similarly for $\delta b$.
\begin{align*}
a^{k+1} - \bar{a} &= tD\laplacian a^{k+1} + tf(a^{k+1}, b^{k+1})\\
a^{k} + \delta a - \bar{a} &= tD\laplacian a^{k} + tD\laplacian \delta a +
 tf(a^k, b^k) + tf_a(a^k, b^{k})\delta a + tf_b(a^k, b^k)\delta b
\end{align*}
where \(f_a\), and \(f_b\) are the derivatives of \(f\) with respect to \(a\)
and \(b\) respectively.  Collecting the terms in $\delta a$ and $\delta b$ on
the LHS,
\begin{align*}
\left(1 - tf_a(a^{k}, b^k)\right)\delta a -tD\laplacian \delta a - 
t f_b(a^k, b^k)\delta b &= \bar{a} - a^{k} + tD\laplacian a^k + tf(a^k, b^k)
\end{align*}
Multiply both sides by $u$, the test function, and integrate by parts setting
the boundary terms to zero due to the boundary conditions (which will be either no flux or periodic).
\begin{align}
 \label{eq:first}
\int_{\Omega}\left(1 - tf_a(a^{k}, b^k)\right)u\delta a\dd{\Omega} +
tD\int_{\Omega} \grad u\cdot \grad \delta a\dd{\Omega} - t\int_{\Omega}
f_b(a^k, b^k)u\delta b\dd{\Omega} =\\
\nonumber \int_{\Omega} u(\bar{a} - a^k)\dd{\Omega} -tD\int_{\Omega}{\grad u
\cdot \grad a^k}\dd{\Omega} + t\int_{\Omega} uf(a^k, b^k)\dd{\Omega}
\end{align}
We will simplify the constraint equation using these linearizations, $a =
a^{k+1} = a^{k} + \delta a$ and $b = b^{k+1} = b^k + \delta b$:
\begin{align*}
	b^{k} + \delta b = \alpha - \frac{1}{S}\int_{\Omega}\left(a^{k}+\delta
	a\right)\dd{\Omega}
\end{align*}
Collecting $\delta a$ and $\delta b$ on the LHS, we get
\begin{align}
 \label{eq:second}
 \left(\frac{1}{S}\int_{\Omega}\delta a\dd{\Omega}\right) + \delta b =
  \alpha - b^k - \frac{1}{S}\int_{\Omega} a^k\dd{\Omega}
\end{align}
The integral equations~\ref{eq:first} and~\ref{eq:second} are the weak forms of
the original PDEs. Our goal is to solve for \(\delta a\), and \(\delta b\). But
this is an infinite-dimensional problem because \(\delta a\) varies continuously
over the surface although \(\delta b\) is uniform. FEM converts the
infinite-dimensional problem to a finite-dimensional problem i.e. a system of
equations. It does this by discretizing the domain of the system of PDEs into a
discrete mesh of simple geometric entities called as the ``finite elements''
like triangles, or rectangles (or tetrahedrons and hexahedrons for
higher-dimensional domains). In our problem, the domain \(\Omega\) is a surface.
The surface is discretized into a mesh of triangles \(\Omega_e\) where
\(e=1,2,\ldots\). \fig\ref{fig:mesh} shows a sample surface domain discretized
into linear triangular elements.
The next step in FEM is to discretize all the field variables in all the
integrands in the above equations into their values at the ``nodes'' of the
finite element. \fig\ref{fig:mesh} shows the nodes of a sample finite element
mesh along with one possible way to index the nodes globally and locally. The
number and location of nodes depends on the type of finite element being used.
A node of the finite element mesh may be shared between multiple finite elements
e.g. in \fig\ref{fig:mesh} global node 1 is the same as local node 1 for
triangle (1) and local node 3 for triangle (2). It is useful to define a
mapping between the global and local indices of the nodes. Let \(E\) be the
total number of elements in the mesh, \(G\) be the total number of nodes in the
finite element mesh, and let \(N\) denote the number of nodes per element. Let
\(I\) denote the global index of a node and \(i^{(e)}\) denote its local index
in the \(e^{th}\) element. We can define a mapping between the global and local
nodes
\begin{align*}
	\Lambda^{(e)}:\{I\}_{I=1}^{I=G} \to \{i^{(e)}\}_{i=1}^{i=N}
\end{align*}
or equivalently
\begin{align}
	\label{eq:global_local_map_1}
	i^{(e)} = \sum_{I=1}^{G}\Lambda^{(e)}_{Ii} I\quad e=1,2,\ldots,E
\end{align}
where \(\Lambda^{(e)}_{Ii}\) denotes an element of an array of zeros and ones defined as
\begin{align}
	\label{eq:global_local_map_2}
	\Lambda^{(e)}_{Ii} = 
	\begin{cases}
		1 & \textrm{if global node \(I\) is the local node \(i\) of element \(e\)}\\
		0 & \textrm{otherwise}
	\end{cases}	
\end{align}
Clearly, \(\Lambda^{(e)}\) is a matrix of size \(N\times G\). As an example, in
reference to \fig\ref{fig:mesh}, the global node index vector is \(\mathbf{I} =
\{1,2,3,4,5,6.7,8,9\}^T\). The local node index vector for triangle (5) is
\(\mathbf{i} = \{5,4,8\}^T\) that gives the global node indices for nodes 1, 2,
and 3 of triangle (5). The mapping matrix \(\Lambda^{(5)}\) as per
equations~\ref{eq:global_local_map_1}, and~\ref{eq:global_local_map_2} is
\begin{align*}
	\Lambda^{(5)} =
	 \begin{bmatrix}
		0 & 0 & 0 & 0 & 1 & 0 & 0 & 0 & 0\\ 
		0 & 0 & 0 & 1 & 0 & 0 & 0 & 0 & 0\\ 
		0 & 0 & 0 & 0 & 0 & 0 & 0 & 1 & 0\\ 
	 \end{bmatrix}.
\end{align*}
We can verify that
\begin{align*}
	\begin{Bmatrix}
		5\\
		4\\
		8
	\end{Bmatrix} &= 
	 \begin{bmatrix}
		0 & 0 & 0 & 0 & 1 & 0 & 0 & 0 & 0\\ 
		0 & 0 & 0 & 1 & 0 & 0 & 0 & 0 & 0\\ 
		0 & 0 & 0 & 0 & 0 & 0 & 0 & 1 & 0\\ 
	 \end{bmatrix}
	 \begin{Bmatrix}
		 1\\
		 2\\
		 3\\
		 4\\
		 5\\
		 6\\
		 7\\
		 8\\
		 9
	 \end{Bmatrix}
\end{align*}
As we noted above that in FEM the field variables are discretized to their nodal
values.  To obtain the value of the field variables at any position other than
the nodes, interpolation is used. The interpolation functions are spatially
varying and their exact form depends on the choice of the finite element. There
is an interpolation function associated with each node of each element. These
are called as the \textit{local interpolation functions}. Let \(\phi^{(e)}_i\)
denote the local interpolation function associated with the \(i^{th}\) node of
the \(e^{th}\) element. These functions satisfy some useful properties like
\begin{align}
	\label{eq:local_props}
	\phi^{(e)}_i(x, y) &\equiv 0\quad \textrm{if } (x,y) \notin \Omega_e\\
	\nonumber \phi^{(e)}_i\left(x^{(e)}_j, y^{(e)}_j\right) &= \delta_{ij}\\
	D^{p}\phi^{(e)}_i\left(x^{(e)}_j, y^{(e)}_j\right) &= \delta_{ij}
\end{align}
where \(\delta_{ij}\) is the Kronecker-delta which is 0 when \(i\neq j\) and 1
when \(i = j\). \(\left(x^{(e)}_j, y^{(e)}_j\right)\) are the coordinates of the
\(j^{th}\) node of the element \(e\). \(D^{p}\) indicates the \(p^{th}\) order
derivative with respect to \(x^{(e)}\), and \(y^{(e)}\) where
\(p\) is less than or equal to the order of \(\phi^{(e)}_i(x, y)\). For example,
a linear triangular finite element (that we have used in our simulations) with
element index \(e\) has three nodes that coincide with its vertices as shown in
\fig\ref{fig:mesh}. If the vertices are located at \(\left(x_1,
y_1\right),\left(x_2, y_2\right)\), and \(\left(x_3,y_3\right)\), the
corresponding local interpolation functions are
\begin{align*}
	\phi^e_1(x, y) = \frac{x \left(y_{2} - y_{3}\right) + x_{2} y_{3} - x_{3} y_{2} - y \left(x_{2} - x_{3}\right)}{x_{1} y_{2} - x_{1} y_{3} - x_{2} y_{1} + x_{2} y_{3} + x_{3} y_{1} - x_{3} y_{2}} \\
	\phi^e_2(x, y) = \frac{- x \left(y_{1} - y_{3}\right) - x_{1} y_{3} + x_{3} y_{1} + y \left(x_{1} - x_{3}\right)}{x_{1} y_{2} - x_{1} y_{3} - x_{2} y_{1} + x_{2} y_{3} + x_{3} y_{1} - x_{3} y_{2}}\\
	\phi^e_3(x, y) = \frac{x \left(y_{1} - y_{2}\right) + x_{1} y_{2} - x_{2} y_{1} - y \left(x_{1} - x_{2}\right)}{x_{1} y_{2} - x_{1} y_{3} - x_{2} y_{1} + x_{2} y_{3} + x_{3} y_{1} - x_{3} y_{2}}
\end{align*}
The local interpolation functions can be ``assembled'' together using the index
mapping shown in equation~\ref{eq:global_local_map_1} into \textit{global
interpolation functions} such that there is one interpolation function
associated with each global node of the finite element mesh. The global
interpolation function associated with the \(I^{th}\) node is
\begin{align}
	\label{eq:global_interpolant}
	\psi_I(x,y) = \bigcup_{e=1}^{E}\sum_{i=1}^{N} \Lambda_{iI}^{(e)}\phi_i^{(e)}(x,y).
\end{align}
Finally,
we are ready to discretize the field variables as interpolations of nodal values
over the entire mesh using the global interpolation functions of
equation~\ref{eq:global_interpolant}. We will use capital letters to denote the
nodal values e.g. \(F_{aI}\) will denote the value of \(f_a(x,y)\) on the
\(I^{th}\) global node. We will use the following substitutions
\begin{align*}
	f_a\left(a(x, y), b\right) &= \sum_{l=1}^G F_{al}\,\psi_l(x, y)\\
	f_b\left(a(x, y), b\right) &= \sum_{m=1}^G F_{bm}\,\psi_m(x, y)\\
    \delta a(x, y) &= \sum_{i=1}^G \delta A_i\,\psi_i(x, y)\\
    \grad \delta a(x, y) &= \sum_{i=1}^G\delta A_i\grad \psi_i(x, y)\\
    \bar{a}(x, y) &= \sum_{n=1}^G\bar{A}_n\psi_n(x, y)\\
    a^k(x, y) &= \sum_{n=1}^GA^k_n\psi_n(x, y)\\
    \grad a^k(x, y) &= \sum_{p=1}^GA^k_p\grad \psi_p(x, y)\\
    f\left(a^k(x,y), b^k\right) &= \sum_{q=1}^GF_q\psi_q(x, y)
\end{align*}
The nodal values \(\delta A_i, i=1,2,\ldots,G\) are the unknowns that we have to
solve for. Therefore, we need \(G\) equations to solve for the \(G\) unknowns.
We generate these equations by setting \(u = \psi_j\) where \(j = 1,
2,\ldots,G\) in equation~\ref{eq:first}. Using the above substitutions and
setting \(u=\psi_j\), equations~\ref{eq:first}, and~\ref{eq:second} can be
written as

\begin{align}
 \label{eq:first_discrete}
\int_{\Omega}\left(1 - t\sum_lF_{al}\psi_l(x,y)\right)\psi_j(x,y)\sum_i\delta
A_i\psi_i(x,y)\dd{\Omega} + tD\int_{\Omega} \grad \psi_j(x,y)\cdot \sum_i \delta
A_i\grad\psi_i(x,y)\dd{\Omega} - t\int_{\Omega} \psi_j(x,y)\sum_mF_{bm}\psi_m(x,y) \delta
b\dd{\Omega} \\\nonumber
= \int_{\Omega} \psi_j(x,y)\left(\sum_n\left(\bar{A}_n - A^k_n\right)\psi_n(x,y)
\right)\dd{\Omega} -tD\int_{\Omega}{\grad \psi_j(x,y) \cdot \sum_pA^k_p\grad
\psi_p(x,y)}\dd{\Omega} + t\int_{\Omega} \psi_j(x,y)\sum_qF_q\psi_q(x,y)\dd{\Omega}
\end{align}
\begin{align}
    \label{eq:second_discrete}
    \frac{1}{S}\int_{\Omega}\sum_i\delta A_i\psi_i(x,y)\dd{\Omega} + \delta b =
    \alpha - b^k - \frac{1}{S}\int_{\Omega} \sum_mA^k_m\psi_m(x,y)\dd{\Omega}
\end{align}
We can rearrange the above equations as
\begin{align}
 \label{eq:first_discrete_arranged}
\sum_i \delta A_i \left[\int_{\Omega}\left(1 -
t\sum_lF_{al}\psi_l(x,y)\right)\psi_i(x,y)\psi_j(x,y) + tD\grad \psi_i(x,y)
\cdot \grad \psi_j(x,y)\dd{\Omega}\right] - \delta b \left(t\int_{\Omega}
\psi_j(x,y)\sum_mF_{bm}\psi_m(x,y) \dd{\Omega}\right) \\
\nonumber = \sum_n\left(\bar{A}_n
- A^k_l\right)\int_{\Omega} \psi_n(x,y)\psi_j(x,y) \dd{\Omega}
-\sum_pA_p^k\left(tD\int_{\Omega}{\grad \psi_p(x,y) \cdot \grad \psi_j(x,y)}
\dd{\Omega}\right) + t\int_{\Omega} \psi_j(x,y)\sum_qF_q\psi_q(x,y)\dd{\Omega}
\end{align}
\begin{align}
    \label{eq:second_discrete_arranged}
    \sum_i \delta A_i \left(\frac{1}{S}\int_{\Omega}\psi_i(x,y)\dd{\Omega}\right) +
    \delta b = \alpha - b^k - \frac{1}{S}\int_{\Omega}
    \sum_mA^{k}_m\psi_m(x,y)\dd{\Omega}
\end{align}
We can write the above equations as 
\begin{align}
    \label{eq:system_of_eqn}
    \sum_iP_{ij}\delta A_i  + Q_j\delta b &= R_j \\
    \nonumber\sum_i S_i \delta A_i  + \delta b &= T
\end{align}
where
\begin{align*}
    P_{ij} &= \int_{\Omega}\left(1 - t\sum_lF_{al}\psi_l(x,y)\right)\psi_i(x,y)\psi_j(x,y)\dd{\Omega}
+ tD\int_{\Omega}\grad \psi_i(x,y) \cdot \grad \psi_j(x,y)\dd{\Omega}\\
    Q_j &= -t\int_{\Omega} \psi_j(x,y)\sum_mF_{bm}\psi_m(x,y) \dd{\Omega}\\
    R_j &= \sum_l\left(\bar{A}_l - A^k_l\right)\int_{\Omega} \psi_l(x,y)\psi_j(x,y) \dd{\Omega}
 -\sum_mA_m^k\left(tD\int_{\Omega}{\grad \psi_m(x,y) \cdot \grad \psi_j(x,y)} \dd{\Omega}\right)
  + t\int_{\Omega} \psi_j(x,y)\sum_qF_q\psi_q(x,y)\dd{\Omega}\\
    S_i &= \frac{1}{S}\int_{\Omega}\psi_i(x,y)\\
    T &= \alpha - b^k - \frac{1}{S}\int_{\Omega} \sum_mA^k_m\psi_m(x,y)\dd{\Omega}
\end{align*}
Equation~\ref{eq:system_of_eqn} can also be written as a block matrix equation
\begin{align*}
\begin{pmatrix}
 [P] & \{Q\} \\[0.25cm]
 \{S\}^T & 1
\end{pmatrix}
\begin{pmatrix}
 \{\delta A\}\\[0.25cm]
 \delta b
\end{pmatrix}
=
\begin{pmatrix}
 \{R\}\\[0.25cm]
 T
\end{pmatrix}
\end{align*}

 All the integrals in the definition of the matrices above will be calculated
 using Gaussian Quadrature over triangles. In general, for solving the above
 system of equations on a 3D surface, the Laplacian operator in the original PDEs
 has to be the Laplace-Beltrami operator. But if we discretize the 3D
 surface using linear triangles we can use the regular Laplacian
 operator~\cite{rognes_2013_automating}. We use FEniCS~\cite{alnaes2015fenics}
 for implementing the finite element method. FEniCS provides various solvers
 for solving the linearized system of equations that we obtained above. We use
 the default LU decomposition solver.

\subsection{Perimeter Minimization Model}

{In our perimeter minimization model of cell polarity, the domain boundary moves to locally minimize the cell's perimeter, while keeping the area constant, i.e. minimizing Equation~\ref{eq:perimin}. This energy $\mathcal{F}$ is a function of the domain perimeter $L$ and its area $A$. To describe the domain dynamics we need to be able to compute the length of a
closed curve embedded in the 3D surface, and the surface area enclosed by the
curve. We will also need the derivatives of the length and the area with respect
to the points parameterizing the domain shape. In the following text, we show how to compute these.} We will derive these results using the first fundamental
form from differential geometry.

 Consider a surface $\Omega$ whose points are defined as
 \begin{equation*}
 \Omega = \left\{ \left(x,\ y,\ h(x, y)\right)\ \rvert\ (x,y)
  \in \mathbb{R}^2\ \mathrm{and}\ h \in \mathbb{R} \right\}.
 \end{equation*}
 We define a piecewise continuous closed curve $\gamma$ embedded in the
 surface $\Omega$. Let its projection in the $x$-$y$ plane be $\Gamma$. We
 will parameterize $\Gamma$ using polar-coordinates centered at a point $(a,
 b)$ in the interior of $\Gamma$ in the $x$-$y$ plane.
 \begin{equation*}
 \Gamma(\theta) = \left(a + r(\theta)\cos\theta,\ b + r(\theta)\sin\theta\right).
 \end{equation*}
 We choose $r(\theta)$ to be a $N$ segment piecewise linear function of $\theta$ such that the
 $k^{\mathrm{th}}$ segment of $r(\theta)$ is
 \begin{equation*}
 r^{(k)}\left(\theta\right) = r_k +
 \frac{r_{k+1} - r_k}{\theta_{k+1} - \theta_k}
 \left(\theta - \theta_k\right)\qquad\mathrm{where}\ k=1,2,...,N.
 \end{equation*}
 Since $\Gamma$ is a closed curve we have an effective periodic boundary
 condition, $r_{N+1} = r_1$ and $\theta_{N+1} = \theta_1$, i.e.
 \begin{equation*}
 r^{(N)}\left(\theta\right) = r_N +
 \frac{r_1 - r_N}{\theta_1 - \theta_N}\left(\theta - \theta_N\right).
 \end{equation*}
 In the above equations we have used
 \begin{equation*}
 r_{k} = r(\theta_k) \qq{where} \theta_k = \frac{2\pi (k-1)}{N}
 \end{equation*}
 The values of \(r^k\)  are the unknown parameters of our optimization problem.
 The coordinates \(\left(a, b\right)\) are calculated as the mean of the \(x\),
 and \(y\) coordinates calculated of from the values of \(r^k\), and
 \(\theta^k\) after every iteration of the gradient-descent algorithm.

 A point $\mathbf{X}$ on the surface $\Omega$ in the interior of $\gamma$ can be
 written as
 \begin{equation*}
 \mathbf{X}(\theta) = \left[x\left(r(\theta),\theta\right),\ y
 \left(r(\theta), \theta\right),\ h(x, y)\right]^{T}
 \end{equation*}
 The coefficients of the first fundamental form can be obtained as
 \begin{align*}
 E &= \pdv{\mathbf{X}}{r}\dotproduct\pdv{\mathbf{X}}{r}\\
 F &= \pdv{\mathbf{X}}{r}\dotproduct\pdv{\mathbf{X}}{\theta}\\
 G &= \pdv{\mathbf{X}}{\theta}\dotproduct\pdv{\mathbf{X}}{\theta}
 \end{align*}

 \subsubsection{Length}

 Length of the $k^{\mathrm{th}}$ segment of the curve $\gamma$ is
 \begin{equation*}
 L^{(k)} = \int_{\theta_k}^{\theta_{k+1}}\,\sqrt{Er'^2 + 2Fr' + G}\,\mathrm{d}\theta
 \qq{where}r' = \pdv{r^{(k)}}{\theta} = \frac{r_{k+1} - r_k}{\theta_{k+1} - \theta_k}.
 \end{equation*}
 Let $f(r, r';\theta) = \sqrt{Er'^2 + 2Fr' + G}$ where $E$, $F$ and $G$ are
 functions of $r$. Then using 1-point Gaussian quadrature we can write
 \begin{equation*}
 L^{(k)} \approx \left(\theta_{k+1} - \theta_k\right)
 f\left({\frac{\theta_{k+1} + \theta_k}{2}}\right)
 \end{equation*}
 It should also be noted that, because we are choosing $r(\theta)$ to be
 piecewise linear, we have
 \begin{equation}
 \label{eq:r}
 r\left({\frac{\theta_{k+1} + \theta_k}{2}}\right) = \frac{r_{k+1} + r_k}{2}.
 \end{equation}
 Using chain-rule and product rule of differentiation,
 \begin{equation*}
 \pdv{L^{(k)}}{r_k} = \left(\theta_{k+1} - \theta_k\right)
 \left(\pdv{f}{r}\pdv{r}{r_k} + \pdv{f}{r'}\pdv{r'}{r_k} \right).
 \end{equation*}
 Therefore,
 \begin{equation*}
 \pdv{L^{(k)}}{r_k} = \left(\frac{\theta_{k+1}-\theta_{k}}{2}\right)\pdv{f}{r} - \pdv{f}{r'}.
 \end{equation*}
 Similarly, we can show that
 \begin{equation*}
 \pdv{L^{(k)}}{r_{k+1}} =\left(\frac{\theta_{k+1}-\theta_{k}}{2}\right)\pdv{f}{r} + \pdv{f}{r'}.
 \end{equation*}

 \subsubsection{Area}

 Area of the $k^{\mathrm{th}}$ triangular slice of the region bounded by
 $\gamma$ can be obtained as
 \begin{equation*}
 A^{(k)} = \int_{\theta_k}^{\theta_{k+1}}\int_0^{r(\theta)}\sqrt{EG - F^2}\,\dd{r}\dd{\theta}
 \end{equation*}
 We will use 1-point Gaussian quadrature along $\theta$ which gives
 \begin{equation*}
 A^{(k)} \approx \left(\theta_{k+1}-\theta_k\right)\int_0^{\frac{r_{k+1}+r_k}{2}}
 \sqrt{EG - F^2}\,\dd{r}
 \end{equation*}
 We will use a 5-point Gaussian quadrature along $r$. Let's use $g \equiv \sqrt{EG-F^2}$.
 \begin{equation*}
 A^{(k)} \approx \left(\theta_{k+1}-\theta_k\right)\left(\frac{r_{k+1}+r_k}{4}\right)
 \sum_i w_i\,g\left(\left(1 + \xi_i\right)\left(\frac{r_{k+1}+r_k}{4}\right)\right)
 \end{equation*}
 where $w_i$ are the Gaussian quadrature weights and $\xi_i$ are the
 corresponding quadrature points.
 We can now write
 \begin{equation*}
 \pdv{A^{(k)}}{r_k} = \left(\frac{\theta_{k+1} - \theta_k}{4}\right)\sum_i\,w_i\,g_i +
 \left(\frac{\theta_{k+1} - \theta_k}{4}\right)\left(r_{k+1} + r_k\right)
 \sum_i\,w_i\,\pdv{g_i}{r}\pdv{r}{r_k}
 \end{equation*}
 We have to evaluate $g_i$ at $r = (1 + \xi_i)(r_{k+1}+r_k)/4$, therefore
 $\partial r/\partial r_k = (1 + \xi_i)/4$,
 \begin{equation*}
 \pdv{A^{(k)}}{r_k} = \left(\frac{\theta_{k+1} - \theta_k}{4}\right)\sum_i\,w_i\,g_i +
 \left(\frac{\theta_{k+1} - \theta_k}{16}\right)
	 \left(r_{k+1} + r_k\right)
	 \sum_i\,(1 + \xi_i)\,w_i\,\pdv{g_i}{r}
 \end{equation*}
 Similarly, it can be shown that
 \begin{equation*}
 \pdv{A^{(k)}}{r_k} = \pdv{A^{(k)}}{r_{k+1}}
 \end{equation*}
Together, these formulas let us compute the domain perimeter and
its area by summing over the areas and lengths in each triangular slice $k$.
Similarly, we can use the derivatives of the area for each slice in order to
evolve Eq. \ref{eq:gradient_descent}  


\begin{thebibliography}{64}%
\makeatletter
\providecommand \@ifxundefined [1]{%
 \@ifx{#1\undefined}
}%
\providecommand \@ifnum [1]{%
 \ifnum #1\expandafter \@firstoftwo
 \else \expandafter \@secondoftwo
 \fi
}%
\providecommand \@ifx [1]{%
 \ifx #1\expandafter \@firstoftwo
 \else \expandafter \@secondoftwo
 \fi
}%
\providecommand \natexlab [1]{#1}%
\providecommand \enquote  [1]{``#1''}%
\providecommand \bibnamefont  [1]{#1}%
\providecommand \bibfnamefont [1]{#1}%
\providecommand \citenamefont [1]{#1}%
\providecommand \href@noop [0]{\@secondoftwo}%
\providecommand \href [0]{\begingroup \@sanitize@url \@href}%
\providecommand \@href[1]{\@@startlink{#1}\@@href}%
\providecommand \@@href[1]{\endgroup#1\@@endlink}%
\providecommand \@sanitize@url [0]{\catcode `\\12\catcode `\$12\catcode
  `\&12\catcode `\#12\catcode `\^12\catcode `\_12\catcode `\%12\relax}%
\providecommand \@@startlink[1]{}%
\providecommand \@@endlink[0]{}%
\providecommand \url  [0]{\begingroup\@sanitize@url \@url }%
\providecommand \@url [1]{\endgroup\@href {#1}{\urlprefix }}%
\providecommand \urlprefix  [0]{URL }%
\providecommand \Eprint [0]{\href }%
\providecommand \doibase [0]{http://dx.doi.org/}%
\providecommand \selectlanguage [0]{\@gobble}%
\providecommand \bibinfo  [0]{\@secondoftwo}%
\providecommand \bibfield  [0]{\@secondoftwo}%
\providecommand \translation [1]{[#1]}%
\providecommand \BibitemOpen [0]{}%
\providecommand \bibitemStop [0]{}%
\providecommand \bibitemNoStop [0]{.\EOS\space}%
\providecommand \EOS [0]{\spacefactor3000\relax}%
\providecommand \BibitemShut  [1]{\csname bibitem#1\endcsname}%
\let\auto@bib@innerbib\@empty
\bibitem [{\citenamefont {Doyle}\ \emph {et~al.}(2013)\citenamefont {Doyle},
  \citenamefont {Petrie}, \citenamefont {Kutys},\ and\ \citenamefont
  {Yamada}}]{doyle_2013_dimensions}%
  \BibitemOpen
  \bibfield  {author} {\bibinfo {author} {\bibfnamefont {A.~D.}\ \bibnamefont
  {Doyle}}, \bibinfo {author} {\bibfnamefont {R.~J.}\ \bibnamefont {Petrie}},
  \bibinfo {author} {\bibfnamefont {M.~L.}\ \bibnamefont {Kutys}}, \ and\
  \bibinfo {author} {\bibfnamefont {K.~M.}\ \bibnamefont {Yamada}},\
  }\href@noop {} {\bibfield  {journal} {\bibinfo  {journal} {Current opinion in
  cell biology}\ }\textbf {\bibinfo {volume} {25}},\ \bibinfo {pages} {642}
  (\bibinfo {year} {2013})},\ \bibinfo {note} {publisher: Elsevier}\BibitemShut
  {NoStop}%
\bibitem [{\citenamefont {Baker}\ and\ \citenamefont
  {Chen}(2012)}]{baker_2012_deconstructing}%
  \BibitemOpen
  \bibfield  {author} {\bibinfo {author} {\bibfnamefont {B.~M.}\ \bibnamefont
  {Baker}}\ and\ \bibinfo {author} {\bibfnamefont {C.~S.}\ \bibnamefont
  {Chen}},\ }\href@noop {} {\bibfield  {journal} {\bibinfo  {journal} {Journal
  of cell science}\ }\textbf {\bibinfo {volume} {125}},\ \bibinfo {pages}
  {3015} (\bibinfo {year} {2012})},\ \bibinfo {note} {publisher: The Company of
  Biologists Ltd}\BibitemShut {NoStop}%
\bibitem [{\citenamefont {Haupt}\ and\ \citenamefont
  {Minc}(2018)}]{haupt_2018_how}%
  \BibitemOpen
  \bibfield  {author} {\bibinfo {author} {\bibfnamefont {A.}~\bibnamefont
  {Haupt}}\ and\ \bibinfo {author} {\bibfnamefont {N.}~\bibnamefont {Minc}},\
  }\href@noop {} {\bibfield  {journal} {\bibinfo  {journal} {Journal of Cell
  Science}\ }\textbf {\bibinfo {volume} {131}} (\bibinfo {year}
  {2018})}\BibitemShut {NoStop}%
\bibitem [{\citenamefont {Simunovic}\ \emph {et~al.}(2016)\citenamefont
  {Simunovic}, \citenamefont {Evergren}, \citenamefont {Golushko},
  \citenamefont {Prévost}, \citenamefont {Renard}, \citenamefont {Johannes},
  \citenamefont {McMahon}, \citenamefont {Lorman}, \citenamefont {Voth},\ and\
  \citenamefont {Bassereau}}]{simunovic_2016_how}%
  \BibitemOpen
  \bibfield  {author} {\bibinfo {author} {\bibfnamefont {M.}~\bibnamefont
  {Simunovic}}, \bibinfo {author} {\bibfnamefont {E.}~\bibnamefont {Evergren}},
  \bibinfo {author} {\bibfnamefont {I.}~\bibnamefont {Golushko}}, \bibinfo
  {author} {\bibfnamefont {C.}~\bibnamefont {Prévost}}, \bibinfo {author}
  {\bibfnamefont {H.-F.}\ \bibnamefont {Renard}}, \bibinfo {author}
  {\bibfnamefont {L.}~\bibnamefont {Johannes}}, \bibinfo {author}
  {\bibfnamefont {H.~T.}\ \bibnamefont {McMahon}}, \bibinfo {author}
  {\bibfnamefont {V.}~\bibnamefont {Lorman}}, \bibinfo {author} {\bibfnamefont
  {G.~A.}\ \bibnamefont {Voth}}, \ and\ \bibinfo {author} {\bibfnamefont
  {P.}~\bibnamefont {Bassereau}},\ }\href@noop {} {\bibfield  {journal}
  {\bibinfo  {journal} {Proceedings of the National Academy of Sciences}\
  }\textbf {\bibinfo {volume} {113}},\ \bibinfo {pages} {11226} (\bibinfo
  {year} {2016})},\ \bibinfo {note} {publisher: National Acad
  Sciences}\BibitemShut {NoStop}%
\bibitem [{\citenamefont {Peter}\ \emph {et~al.}(2004)\citenamefont {Peter},
  \citenamefont {Kent}, \citenamefont {Mills}, \citenamefont {Vallis},
  \citenamefont {Butler}, \citenamefont {Evans},\ and\ \citenamefont
  {McMahon}}]{peter_2004_bar}%
  \BibitemOpen
  \bibfield  {author} {\bibinfo {author} {\bibfnamefont {B.~J.}\ \bibnamefont
  {Peter}}, \bibinfo {author} {\bibfnamefont {H.~M.}\ \bibnamefont {Kent}},
  \bibinfo {author} {\bibfnamefont {I.~G.}\ \bibnamefont {Mills}}, \bibinfo
  {author} {\bibfnamefont {Y.}~\bibnamefont {Vallis}}, \bibinfo {author}
  {\bibfnamefont {P.~J.~G.}\ \bibnamefont {Butler}}, \bibinfo {author}
  {\bibfnamefont {P.~R.}\ \bibnamefont {Evans}}, \ and\ \bibinfo {author}
  {\bibfnamefont {H.~T.}\ \bibnamefont {McMahon}},\ }\href@noop {} {\bibfield
  {journal} {\bibinfo  {journal} {Science}\ }\textbf {\bibinfo {volume}
  {303}},\ \bibinfo {pages} {495} (\bibinfo {year} {2004})},\ \bibinfo {note}
  {publisher: American Association for the Advancement of Science}\BibitemShut
  {NoStop}%
\bibitem [{\citenamefont {Bigay}\ \emph {et~al.}(2005)\citenamefont {Bigay},
  \citenamefont {Casella}, \citenamefont {Drin}, \citenamefont {Mesmin},\ and\
  \citenamefont {Antonny}}]{bigay_2005_arfgap1}%
  \BibitemOpen
  \bibfield  {author} {\bibinfo {author} {\bibfnamefont {J.}~\bibnamefont
  {Bigay}}, \bibinfo {author} {\bibfnamefont {J.-F.}\ \bibnamefont {Casella}},
  \bibinfo {author} {\bibfnamefont {G.}~\bibnamefont {Drin}}, \bibinfo {author}
  {\bibfnamefont {B.}~\bibnamefont {Mesmin}}, \ and\ \bibinfo {author}
  {\bibfnamefont {B.}~\bibnamefont {Antonny}},\ }\href@noop {} {\bibfield
  {journal} {\bibinfo  {journal} {The EMBO journal}\ }\textbf {\bibinfo
  {volume} {24}},\ \bibinfo {pages} {2244} (\bibinfo {year} {2005})},\ \bibinfo
  {note} {publisher: John Wiley \& Sons, Ltd Chichester, UK}\BibitemShut
  {NoStop}%
\bibitem [{\citenamefont {Drin}\ \emph {et~al.}(2007)\citenamefont {Drin},
  \citenamefont {Casella}, \citenamefont {Gautier}, \citenamefont {Boehmer},
  \citenamefont {Schwartz},\ and\ \citenamefont {Antonny}}]{drin_2007_general}%
  \BibitemOpen
  \bibfield  {author} {\bibinfo {author} {\bibfnamefont {G.}~\bibnamefont
  {Drin}}, \bibinfo {author} {\bibfnamefont {J.-F.}\ \bibnamefont {Casella}},
  \bibinfo {author} {\bibfnamefont {R.}~\bibnamefont {Gautier}}, \bibinfo
  {author} {\bibfnamefont {T.}~\bibnamefont {Boehmer}}, \bibinfo {author}
  {\bibfnamefont {T.~U.}\ \bibnamefont {Schwartz}}, \ and\ \bibinfo {author}
  {\bibfnamefont {B.}~\bibnamefont {Antonny}},\ }\href@noop {} {\bibfield
  {journal} {\bibinfo  {journal} {Nature structural \& molecular biology}\
  }\textbf {\bibinfo {volume} {14}},\ \bibinfo {pages} {138} (\bibinfo {year}
  {2007})},\ \bibinfo {note} {publisher: Nature Publishing Group}\BibitemShut
  {NoStop}%
\bibitem [{\citenamefont {Vanni}\ \emph {et~al.}(2014)\citenamefont {Vanni},
  \citenamefont {Hirose}, \citenamefont {Barelli}, \citenamefont {Antonny},\
  and\ \citenamefont {Gautier}}]{vanni_2014_sub}%
  \BibitemOpen
  \bibfield  {author} {\bibinfo {author} {\bibfnamefont {S.}~\bibnamefont
  {Vanni}}, \bibinfo {author} {\bibfnamefont {H.}~\bibnamefont {Hirose}},
  \bibinfo {author} {\bibfnamefont {H.}~\bibnamefont {Barelli}}, \bibinfo
  {author} {\bibfnamefont {B.}~\bibnamefont {Antonny}}, \ and\ \bibinfo
  {author} {\bibfnamefont {R.}~\bibnamefont {Gautier}},\ }\href@noop {}
  {\bibfield  {journal} {\bibinfo  {journal} {Nature communications}\ }\textbf
  {\bibinfo {volume} {5}},\ \bibinfo {pages} {1} (\bibinfo {year} {2014})},\
  \bibinfo {note} {publisher: Nature Publishing Group}\BibitemShut {NoStop}%
\bibitem [{\citenamefont {Pranke}\ \emph {et~al.}(2011)\citenamefont {Pranke},
  \citenamefont {Morello}, \citenamefont {Bigay}, \citenamefont {Gibson},
  \citenamefont {Verbavatz}, \citenamefont {Antonny},\ and\ \citenamefont
  {Jackson}}]{pranke_2011_alpha}%
  \BibitemOpen
  \bibfield  {author} {\bibinfo {author} {\bibfnamefont {I.~M.}\ \bibnamefont
  {Pranke}}, \bibinfo {author} {\bibfnamefont {V.}~\bibnamefont {Morello}},
  \bibinfo {author} {\bibfnamefont {J.}~\bibnamefont {Bigay}}, \bibinfo
  {author} {\bibfnamefont {K.}~\bibnamefont {Gibson}}, \bibinfo {author}
  {\bibfnamefont {J.-M.}\ \bibnamefont {Verbavatz}}, \bibinfo {author}
  {\bibfnamefont {B.}~\bibnamefont {Antonny}}, \ and\ \bibinfo {author}
  {\bibfnamefont {C.~L.}\ \bibnamefont {Jackson}},\ }\href@noop {} {\bibfield
  {journal} {\bibinfo  {journal} {Journal of Cell Biology}\ }\textbf {\bibinfo
  {volume} {194}},\ \bibinfo {pages} {89} (\bibinfo {year} {2011})},\ \bibinfo
  {note} {publisher: The Rockefeller University Press}\BibitemShut {NoStop}%
\bibitem [{\citenamefont {Levin}\ \emph {et~al.}(1993)\citenamefont {Levin},
  \citenamefont {Fan}, \citenamefont {Ricca}, \citenamefont {Driks},
  \citenamefont {Losick},\ and\ \citenamefont
  {Cutting}}]{levin_1993_unusually}%
  \BibitemOpen
  \bibfield  {author} {\bibinfo {author} {\bibfnamefont {P.~A.}\ \bibnamefont
  {Levin}}, \bibinfo {author} {\bibfnamefont {N.}~\bibnamefont {Fan}}, \bibinfo
  {author} {\bibfnamefont {E.}~\bibnamefont {Ricca}}, \bibinfo {author}
  {\bibfnamefont {A.}~\bibnamefont {Driks}}, \bibinfo {author} {\bibfnamefont
  {R.}~\bibnamefont {Losick}}, \ and\ \bibinfo {author} {\bibfnamefont
  {S.}~\bibnamefont {Cutting}},\ }\href@noop {} {\bibfield  {journal} {\bibinfo
   {journal} {Molecular microbiology}\ }\textbf {\bibinfo {volume} {9}},\
  \bibinfo {pages} {761} (\bibinfo {year} {1993})},\ \bibinfo {note}
  {publisher: Wiley Online Library}\BibitemShut {NoStop}%
\bibitem [{\citenamefont {Mostowy}\ and\ \citenamefont
  {Cossart}(2012)}]{mostowy_2012_septins}%
  \BibitemOpen
  \bibfield  {author} {\bibinfo {author} {\bibfnamefont {S.}~\bibnamefont
  {Mostowy}}\ and\ \bibinfo {author} {\bibfnamefont {P.}~\bibnamefont
  {Cossart}},\ }\href@noop {} {\bibfield  {journal} {\bibinfo  {journal}
  {Nature reviews Molecular cell biology}\ }\textbf {\bibinfo {volume} {13}},\
  \bibinfo {pages} {183} (\bibinfo {year} {2012})},\ \bibinfo {note}
  {publisher: Nature Publishing Group}\BibitemShut {NoStop}%
\bibitem [{\citenamefont {Meyers}\ \emph {et~al.}(2006)\citenamefont {Meyers},
  \citenamefont {Craig},\ and\ \citenamefont {Odde}}]{meyers_2006_potential}%
  \BibitemOpen
  \bibfield  {author} {\bibinfo {author} {\bibfnamefont {J.}~\bibnamefont
  {Meyers}}, \bibinfo {author} {\bibfnamefont {J.}~\bibnamefont {Craig}}, \
  and\ \bibinfo {author} {\bibfnamefont {D.~J.}\ \bibnamefont {Odde}},\
  }\href@noop {} {\bibfield  {journal} {\bibinfo  {journal} {Current Biology}\
  }\textbf {\bibinfo {volume} {16}},\ \bibinfo {pages} {1685} (\bibinfo {year}
  {2006})}\BibitemShut {NoStop}%
\bibitem [{\citenamefont {Camley}\ \emph {et~al.}(2017)\citenamefont {Camley},
  \citenamefont {Zhao}, \citenamefont {Li}, \citenamefont {Levine},\ and\
  \citenamefont {Rappel}}]{camley_2017_crawling}%
  \BibitemOpen
  \bibfield  {author} {\bibinfo {author} {\bibfnamefont {B.~A.}\ \bibnamefont
  {Camley}}, \bibinfo {author} {\bibfnamefont {Y.}~\bibnamefont {Zhao}},
  \bibinfo {author} {\bibfnamefont {B.}~\bibnamefont {Li}}, \bibinfo {author}
  {\bibfnamefont {H.}~\bibnamefont {Levine}}, \ and\ \bibinfo {author}
  {\bibfnamefont {W.-J.}\ \bibnamefont {Rappel}},\ }\href@noop {} {\bibfield
  {journal} {\bibinfo  {journal} {Physical Review E}\ }\textbf {\bibinfo
  {volume} {95}},\ \bibinfo {pages} {012401} (\bibinfo {year}
  {2017})}\BibitemShut {NoStop}%
\bibitem [{\citenamefont {Cusseddu}\ \emph {et~al.}(2019)\citenamefont
  {Cusseddu}, \citenamefont {Edelstein-Keshet}, \citenamefont {Mackenzie},
  \citenamefont {Portet},\ and\ \citenamefont
  {Madzvamuse}}]{cusseddu_2019_coupled}%
  \BibitemOpen
  \bibfield  {author} {\bibinfo {author} {\bibfnamefont {D.}~\bibnamefont
  {Cusseddu}}, \bibinfo {author} {\bibfnamefont {L.}~\bibnamefont
  {Edelstein-Keshet}}, \bibinfo {author} {\bibfnamefont {J.}~\bibnamefont
  {Mackenzie}}, \bibinfo {author} {\bibfnamefont {S.}~\bibnamefont {Portet}}, \
  and\ \bibinfo {author} {\bibfnamefont {A.}~\bibnamefont {Madzvamuse}},\
  }\href {\doibase 10.1016/j.jtbi.2018.09.008} {\bibfield  {journal} {\bibinfo
  {journal} {Journal of Theoretical Biology}\ }\textbf {\bibinfo {volume}
  {481}},\ \bibinfo {pages} {119} (\bibinfo {year} {2019})}\BibitemShut
  {NoStop}%
\bibitem [{\citenamefont {Rangamani}\ \emph {et~al.}(2013)\citenamefont
  {Rangamani}, \citenamefont {Lipshtat}, \citenamefont {Azeloglu},
  \citenamefont {Calizo}, \citenamefont {Hu}, \citenamefont {Ghassemi},
  \citenamefont {Hone}, \citenamefont {Scarlata}, \citenamefont {Neves},\ and\
  \citenamefont {Iyengar}}]{rangamani_2013_decoding}%
  \BibitemOpen
  \bibfield  {author} {\bibinfo {author} {\bibfnamefont {P.}~\bibnamefont
  {Rangamani}}, \bibinfo {author} {\bibfnamefont {A.}~\bibnamefont {Lipshtat}},
  \bibinfo {author} {\bibfnamefont {E.~U.}\ \bibnamefont {Azeloglu}}, \bibinfo
  {author} {\bibfnamefont {R.~C.}\ \bibnamefont {Calizo}}, \bibinfo {author}
  {\bibfnamefont {M.}~\bibnamefont {Hu}}, \bibinfo {author} {\bibfnamefont
  {S.}~\bibnamefont {Ghassemi}}, \bibinfo {author} {\bibfnamefont
  {J.}~\bibnamefont {Hone}}, \bibinfo {author} {\bibfnamefont {S.}~\bibnamefont
  {Scarlata}}, \bibinfo {author} {\bibfnamefont {S.~R.}\ \bibnamefont {Neves}},
  \ and\ \bibinfo {author} {\bibfnamefont {R.}~\bibnamefont {Iyengar}},\
  }\href@noop {} {\bibfield  {journal} {\bibinfo  {journal} {Cell}\ }\textbf
  {\bibinfo {volume} {154}},\ \bibinfo {pages} {1356} (\bibinfo {year}
  {2013})}\BibitemShut {NoStop}%
\bibitem [{\citenamefont {Ramirez}\ \emph {et~al.}(2015)\citenamefont
  {Ramirez}, \citenamefont {Raghavachari},\ and\ \citenamefont
  {Lew}}]{ramirez2015dendritic}%
  \BibitemOpen
  \bibfield  {author} {\bibinfo {author} {\bibfnamefont {S.~A.}\ \bibnamefont
  {Ramirez}}, \bibinfo {author} {\bibfnamefont {S.}~\bibnamefont
  {Raghavachari}}, \ and\ \bibinfo {author} {\bibfnamefont {D.~J.}\
  \bibnamefont {Lew}},\ }\href@noop {} {\bibfield  {journal} {\bibinfo
  {journal} {Molecular biology of the cell}\ }\textbf {\bibinfo {volume}
  {26}},\ \bibinfo {pages} {4171} (\bibinfo {year} {2015})}\BibitemShut
  {NoStop}%
\bibitem [{\citenamefont {Frank}\ \emph {et~al.}(2019)\citenamefont {Frank},
  \citenamefont {Guven}, \citenamefont {Kardar},\ and\ \citenamefont
  {Shackleton}}]{frank2019pinning}%
  \BibitemOpen
  \bibfield  {author} {\bibinfo {author} {\bibfnamefont {J.~R.}\ \bibnamefont
  {Frank}}, \bibinfo {author} {\bibfnamefont {J.}~\bibnamefont {Guven}},
  \bibinfo {author} {\bibfnamefont {M.}~\bibnamefont {Kardar}}, \ and\ \bibinfo
  {author} {\bibfnamefont {H.}~\bibnamefont {Shackleton}},\ }\href@noop {}
  {\bibfield  {journal} {\bibinfo  {journal} {EPL (Europhysics Letters)}\
  }\textbf {\bibinfo {volume} {127}},\ \bibinfo {pages} {48001} (\bibinfo
  {year} {2019})}\BibitemShut {NoStop}%
\bibitem [{\citenamefont {Spill}\ \emph {et~al.}(2016)\citenamefont {Spill},
  \citenamefont {Andasari}, \citenamefont {Mak}, \citenamefont {Kamm},\ and\
  \citenamefont {Zaman}}]{spill2016effects}%
  \BibitemOpen
  \bibfield  {author} {\bibinfo {author} {\bibfnamefont {F.}~\bibnamefont
  {Spill}}, \bibinfo {author} {\bibfnamefont {V.}~\bibnamefont {Andasari}},
  \bibinfo {author} {\bibfnamefont {M.}~\bibnamefont {Mak}}, \bibinfo {author}
  {\bibfnamefont {R.~D.}\ \bibnamefont {Kamm}}, \ and\ \bibinfo {author}
  {\bibfnamefont {M.~H.}\ \bibnamefont {Zaman}},\ }\href@noop {} {\bibfield
  {journal} {\bibinfo  {journal} {Physical Biology}\ }\textbf {\bibinfo
  {volume} {13}},\ \bibinfo {pages} {036008} (\bibinfo {year}
  {2016})}\BibitemShut {NoStop}%
\bibitem [{\citenamefont {Eroum{\'e}}\ \emph {et~al.}(2021)\citenamefont
  {Eroum{\'e}}, \citenamefont {Vasilevich}, \citenamefont {Vermeulen},
  \citenamefont {de~Boer},\ and\ \citenamefont
  {Carlier}}]{eroume2021influence}%
  \BibitemOpen
  \bibfield  {author} {\bibinfo {author} {\bibfnamefont {K.}~\bibnamefont
  {Eroum{\'e}}}, \bibinfo {author} {\bibfnamefont {A.}~\bibnamefont
  {Vasilevich}}, \bibinfo {author} {\bibfnamefont {S.}~\bibnamefont
  {Vermeulen}}, \bibinfo {author} {\bibfnamefont {J.}~\bibnamefont {de~Boer}},
  \ and\ \bibinfo {author} {\bibfnamefont {A.}~\bibnamefont {Carlier}},\
  }\href@noop {} {\bibfield  {journal} {\bibinfo  {journal} {PLOS ONE}\
  }\textbf {\bibinfo {volume} {16}},\ \bibinfo {pages} {e0248293} (\bibinfo
  {year} {2021})}\BibitemShut {NoStop}%
\bibitem [{\citenamefont {Elliott}\ \emph {et~al.}(2015)\citenamefont
  {Elliott}, \citenamefont {Fischer}, \citenamefont {Myers}, \citenamefont
  {Desai}, \citenamefont {Gao}, \citenamefont {Chen}, \citenamefont
  {Adelstein}, \citenamefont {Waterman},\ and\ \citenamefont
  {Danuser}}]{elliott_2015_myosin}%
  \BibitemOpen
  \bibfield  {author} {\bibinfo {author} {\bibfnamefont {H.}~\bibnamefont
  {Elliott}}, \bibinfo {author} {\bibfnamefont {R.~S.}\ \bibnamefont
  {Fischer}}, \bibinfo {author} {\bibfnamefont {K.~A.}\ \bibnamefont {Myers}},
  \bibinfo {author} {\bibfnamefont {R.~A.}\ \bibnamefont {Desai}}, \bibinfo
  {author} {\bibfnamefont {L.}~\bibnamefont {Gao}}, \bibinfo {author}
  {\bibfnamefont {C.~S.}\ \bibnamefont {Chen}}, \bibinfo {author}
  {\bibfnamefont {R.~S.}\ \bibnamefont {Adelstein}}, \bibinfo {author}
  {\bibfnamefont {C.~M.}\ \bibnamefont {Waterman}}, \ and\ \bibinfo {author}
  {\bibfnamefont {G.}~\bibnamefont {Danuser}},\ }\href {\doibase
  10.1038/ncb3092} {\bibfield  {journal} {\bibinfo  {journal} {Nature Cell
  Biology}\ }\textbf {\bibinfo {volume} {17}},\ \bibinfo {pages} {137}
  (\bibinfo {year} {2015})},\ \bibinfo {note} {publisher: Nature Publishing
  Group}\BibitemShut {NoStop}%
\bibitem [{\citenamefont {Driscoll}\ \emph {et~al.}(2019)\citenamefont
  {Driscoll}, \citenamefont {Welf}, \citenamefont {Jamieson}, \citenamefont
  {Dean}, \citenamefont {Isogai}, \citenamefont {Fiolka},\ and\ \citenamefont
  {Danuser}}]{driscoll_2019_robust}%
  \BibitemOpen
  \bibfield  {author} {\bibinfo {author} {\bibfnamefont {M.~K.}\ \bibnamefont
  {Driscoll}}, \bibinfo {author} {\bibfnamefont {E.~S.}\ \bibnamefont {Welf}},
  \bibinfo {author} {\bibfnamefont {A.~R.}\ \bibnamefont {Jamieson}}, \bibinfo
  {author} {\bibfnamefont {K.~M.}\ \bibnamefont {Dean}}, \bibinfo {author}
  {\bibfnamefont {T.}~\bibnamefont {Isogai}}, \bibinfo {author} {\bibfnamefont
  {R.}~\bibnamefont {Fiolka}}, \ and\ \bibinfo {author} {\bibfnamefont
  {G.}~\bibnamefont {Danuser}},\ }\href {\doibase 10.1038/s41592-019-0539-z}
  {\bibfield  {journal} {\bibinfo  {journal} {Nature Methods}\ }\textbf
  {\bibinfo {volume} {16}},\ \bibinfo {pages} {1037} (\bibinfo {year}
  {2019})}\BibitemShut {NoStop}%
\bibitem [{\citenamefont {Wigbers}\ \emph {et~al.}(2021)\citenamefont
  {Wigbers}, \citenamefont {Tan}, \citenamefont {Brauns}, \citenamefont {Liu},
  \citenamefont {Swartz}, \citenamefont {Frey},\ and\ \citenamefont
  {Fakhri}}]{wigbers_2021_hierarchy}%
  \BibitemOpen
  \bibfield  {author} {\bibinfo {author} {\bibfnamefont {M.~C.}\ \bibnamefont
  {Wigbers}}, \bibinfo {author} {\bibfnamefont {T.~H.}\ \bibnamefont {Tan}},
  \bibinfo {author} {\bibfnamefont {F.}~\bibnamefont {Brauns}}, \bibinfo
  {author} {\bibfnamefont {J.}~\bibnamefont {Liu}}, \bibinfo {author}
  {\bibfnamefont {S.~Z.}\ \bibnamefont {Swartz}}, \bibinfo {author}
  {\bibfnamefont {E.}~\bibnamefont {Frey}}, \ and\ \bibinfo {author}
  {\bibfnamefont {N.}~\bibnamefont {Fakhri}},\ }\href@noop {} {\bibfield
  {journal} {\bibinfo  {journal} {Nature Physics}\ }\textbf {\bibinfo {volume}
  {17}},\ \bibinfo {pages} {578} (\bibinfo {year} {2021})}\BibitemShut
  {NoStop}%
\bibitem [{\citenamefont {Bonazzi}\ \emph {et~al.}(2015)\citenamefont
  {Bonazzi}, \citenamefont {Haupt}, \citenamefont {Tanimoto}, \citenamefont
  {Delacour}, \citenamefont {Salort},\ and\ \citenamefont
  {Minc}}]{bonazzi2015actin}%
  \BibitemOpen
  \bibfield  {author} {\bibinfo {author} {\bibfnamefont {D.}~\bibnamefont
  {Bonazzi}}, \bibinfo {author} {\bibfnamefont {A.}~\bibnamefont {Haupt}},
  \bibinfo {author} {\bibfnamefont {H.}~\bibnamefont {Tanimoto}}, \bibinfo
  {author} {\bibfnamefont {D.}~\bibnamefont {Delacour}}, \bibinfo {author}
  {\bibfnamefont {D.}~\bibnamefont {Salort}}, \ and\ \bibinfo {author}
  {\bibfnamefont {N.}~\bibnamefont {Minc}},\ }\href@noop {} {\bibfield
  {journal} {\bibinfo  {journal} {Current Biology}\ }\textbf {\bibinfo {volume}
  {25}},\ \bibinfo {pages} {2677} (\bibinfo {year} {2015})}\BibitemShut
  {NoStop}%
\bibitem [{\citenamefont {Schweizer}\ \emph {et~al.}(2012)\citenamefont
  {Schweizer}, \citenamefont {Loose}, \citenamefont {Bonny}, \citenamefont
  {Kruse}, \citenamefont {M{\"o}nch},\ and\ \citenamefont
  {Schwille}}]{schweizer_2012_geometry}%
  \BibitemOpen
  \bibfield  {author} {\bibinfo {author} {\bibfnamefont {J.}~\bibnamefont
  {Schweizer}}, \bibinfo {author} {\bibfnamefont {M.}~\bibnamefont {Loose}},
  \bibinfo {author} {\bibfnamefont {M.}~\bibnamefont {Bonny}}, \bibinfo
  {author} {\bibfnamefont {K.}~\bibnamefont {Kruse}}, \bibinfo {author}
  {\bibfnamefont {I.}~\bibnamefont {M{\"o}nch}}, \ and\ \bibinfo {author}
  {\bibfnamefont {P.}~\bibnamefont {Schwille}},\ }\href@noop {} {\bibfield
  {journal} {\bibinfo  {journal} {Proceedings of the National Academy of
  Sciences}\ }\textbf {\bibinfo {volume} {109}},\ \bibinfo {pages} {15283}
  (\bibinfo {year} {2012})}\BibitemShut {NoStop}%
\bibitem [{\citenamefont {Halatek}\ and\ \citenamefont
  {Frey}(2014)}]{halatek_2014_effective}%
  \BibitemOpen
  \bibfield  {author} {\bibinfo {author} {\bibfnamefont {J.}~\bibnamefont
  {Halatek}}\ and\ \bibinfo {author} {\bibfnamefont {E.}~\bibnamefont {Frey}},\
  }\href@noop {} {\bibfield  {journal} {\bibinfo  {journal} {Proceedings of the
  National Academy of Sciences}\ }\textbf {\bibinfo {volume} {111}},\ \bibinfo
  {pages} {E1817} (\bibinfo {year} {2014})}\BibitemShut {NoStop}%
\bibitem [{\citenamefont {Allen}\ \emph {et~al.}(2020)\citenamefont {Allen},
  \citenamefont {Lee}, \citenamefont {Barnhart}, \citenamefont {Tsuchida},
  \citenamefont {Wilson}, \citenamefont {Gutierrez}, \citenamefont {Groisman},
  \citenamefont {Theriot},\ and\ \citenamefont {Mogilner}}]{allen_2020_cell}%
  \BibitemOpen
  \bibfield  {author} {\bibinfo {author} {\bibfnamefont {G.~M.}\ \bibnamefont
  {Allen}}, \bibinfo {author} {\bibfnamefont {K.~C.}\ \bibnamefont {Lee}},
  \bibinfo {author} {\bibfnamefont {E.~L.}\ \bibnamefont {Barnhart}}, \bibinfo
  {author} {\bibfnamefont {M.~A.}\ \bibnamefont {Tsuchida}}, \bibinfo {author}
  {\bibfnamefont {C.~A.}\ \bibnamefont {Wilson}}, \bibinfo {author}
  {\bibfnamefont {E.}~\bibnamefont {Gutierrez}}, \bibinfo {author}
  {\bibfnamefont {A.}~\bibnamefont {Groisman}}, \bibinfo {author}
  {\bibfnamefont {J.~A.}\ \bibnamefont {Theriot}}, \ and\ \bibinfo {author}
  {\bibfnamefont {A.}~\bibnamefont {Mogilner}},\ }\href@noop {} {\bibfield
  {journal} {\bibinfo  {journal} {Cell Systems}\ }\textbf {\bibinfo {volume}
  {11}},\ \bibinfo {pages} {286} (\bibinfo {year} {2020})}\BibitemShut
  {NoStop}%
\bibitem [{\citenamefont {Camley}\ \emph {et~al.}(2013)\citenamefont {Camley},
  \citenamefont {Zhao}, \citenamefont {Li}, \citenamefont {Levine},\ and\
  \citenamefont {Rappel}}]{camley_2013_periodic}%
  \BibitemOpen
  \bibfield  {author} {\bibinfo {author} {\bibfnamefont {B.~A.}\ \bibnamefont
  {Camley}}, \bibinfo {author} {\bibfnamefont {Y.}~\bibnamefont {Zhao}},
  \bibinfo {author} {\bibfnamefont {B.}~\bibnamefont {Li}}, \bibinfo {author}
  {\bibfnamefont {H.}~\bibnamefont {Levine}}, \ and\ \bibinfo {author}
  {\bibfnamefont {W.-J.}\ \bibnamefont {Rappel}},\ }\href@noop {} {\bibfield
  {journal} {\bibinfo  {journal} {Physical Review Letters}\ }\textbf {\bibinfo
  {volume} {111}},\ \bibinfo {pages} {158102} (\bibinfo {year}
  {2013})}\BibitemShut {NoStop}%
\bibitem [{\citenamefont {Hubatsch}\ \emph {et~al.}(2019)\citenamefont
  {Hubatsch}, \citenamefont {Peglion}, \citenamefont {Reich}, \citenamefont
  {Rodrigues}, \citenamefont {Hirani}, \citenamefont {Illukkumbura},\ and\
  \citenamefont {Goehring}}]{hubatsch_2019_cell}%
  \BibitemOpen
  \bibfield  {author} {\bibinfo {author} {\bibfnamefont {L.}~\bibnamefont
  {Hubatsch}}, \bibinfo {author} {\bibfnamefont {F.}~\bibnamefont {Peglion}},
  \bibinfo {author} {\bibfnamefont {J.~D.}\ \bibnamefont {Reich}}, \bibinfo
  {author} {\bibfnamefont {N.~T.~L.}\ \bibnamefont {Rodrigues}}, \bibinfo
  {author} {\bibfnamefont {N.}~\bibnamefont {Hirani}}, \bibinfo {author}
  {\bibfnamefont {R.}~\bibnamefont {Illukkumbura}}, \ and\ \bibinfo {author}
  {\bibfnamefont {N.~W.}\ \bibnamefont {Goehring}},\ }\href {\doibase
  10.1038/s41567-019-0601-x} {\bibfield  {journal} {\bibinfo  {journal} {Nature
  Physics}\ }\textbf {\bibinfo {volume} {15}},\ \bibinfo {pages} {1078}
  (\bibinfo {year} {2019})},\ \bibinfo {note} {publisher: Nature Publishing
  Group}\BibitemShut {NoStop}%
\bibitem [{\citenamefont {Mittasch}\ \emph {et~al.}(2018)\citenamefont
  {Mittasch}, \citenamefont {Gross}, \citenamefont {Nestler}, \citenamefont
  {Fritsch}, \citenamefont {Iserman}, \citenamefont {Kar}, \citenamefont
  {Munder}, \citenamefont {Voigt}, \citenamefont {Alberti}, \citenamefont
  {Grill},\ and\ \citenamefont {Kreysing}}]{mittasch_2018_non}%
  \BibitemOpen
  \bibfield  {author} {\bibinfo {author} {\bibfnamefont {M.}~\bibnamefont
  {Mittasch}}, \bibinfo {author} {\bibfnamefont {P.}~\bibnamefont {Gross}},
  \bibinfo {author} {\bibfnamefont {M.}~\bibnamefont {Nestler}}, \bibinfo
  {author} {\bibfnamefont {A.~W.}\ \bibnamefont {Fritsch}}, \bibinfo {author}
  {\bibfnamefont {C.}~\bibnamefont {Iserman}}, \bibinfo {author} {\bibfnamefont
  {M.}~\bibnamefont {Kar}}, \bibinfo {author} {\bibfnamefont {M.}~\bibnamefont
  {Munder}}, \bibinfo {author} {\bibfnamefont {A.}~\bibnamefont {Voigt}},
  \bibinfo {author} {\bibfnamefont {S.}~\bibnamefont {Alberti}}, \bibinfo
  {author} {\bibfnamefont {S.~W.}\ \bibnamefont {Grill}}, \ and\ \bibinfo
  {author} {\bibfnamefont {M.}~\bibnamefont {Kreysing}},\ }\href {\doibase
  10.1038/s41556-017-0032-9} {\bibfield  {journal} {\bibinfo  {journal} {Nature
  Cell Biology}\ }\textbf {\bibinfo {volume} {20}},\ \bibinfo {pages} {344}
  (\bibinfo {year} {2018})},\ \bibinfo {note} {publisher: Nature Publishing
  Group}\BibitemShut {NoStop}%
\bibitem [{\citenamefont {Murray}(2001)}]{murray2001mathematical}%
  \BibitemOpen
  \bibfield  {author} {\bibinfo {author} {\bibfnamefont {J.~D.}\ \bibnamefont
  {Murray}},\ }\href@noop {} {\emph {\bibinfo {title} {Mathematical Biology II:
  spatial models and biomedical applications}}},\ Vol.~\bibinfo {volume} {3}\
  (\bibinfo  {publisher} {Springer New York},\ \bibinfo {year}
  {2001})\BibitemShut {NoStop}%
\bibitem [{\citenamefont {Mori}\ \emph {et~al.}(2008)\citenamefont {Mori},
  \citenamefont {Jilkine},\ and\ \citenamefont
  {Edelstein-Keshet}}]{mori_2008_wave}%
  \BibitemOpen
  \bibfield  {author} {\bibinfo {author} {\bibfnamefont {Y.}~\bibnamefont
  {Mori}}, \bibinfo {author} {\bibfnamefont {A.}~\bibnamefont {Jilkine}}, \
  and\ \bibinfo {author} {\bibfnamefont {L.}~\bibnamefont {Edelstein-Keshet}},\
  }\href {\doibase 10.1529/biophysj.107.120824} {\bibfield  {journal} {\bibinfo
   {journal} {Biophysical Journal}\ }\textbf {\bibinfo {volume} {94}},\
  \bibinfo {pages} {3684} (\bibinfo {year} {2008})},\ \bibinfo {note}
  {publisher: Elsevier}\BibitemShut {NoStop}%
\bibitem [{\citenamefont {Goehring}\ \emph
  {et~al.}(2011{\natexlab{a}})\citenamefont {Goehring}, \citenamefont {Trong},
  \citenamefont {Bois}, \citenamefont {Chowdhury}, \citenamefont {Nicola},
  \citenamefont {Hyman},\ and\ \citenamefont
  {Grill}}]{goehring2011polarization}%
  \BibitemOpen
  \bibfield  {author} {\bibinfo {author} {\bibfnamefont {N.~W.}\ \bibnamefont
  {Goehring}}, \bibinfo {author} {\bibfnamefont {P.~K.}\ \bibnamefont {Trong}},
  \bibinfo {author} {\bibfnamefont {J.~S.}\ \bibnamefont {Bois}}, \bibinfo
  {author} {\bibfnamefont {D.}~\bibnamefont {Chowdhury}}, \bibinfo {author}
  {\bibfnamefont {E.~M.}\ \bibnamefont {Nicola}}, \bibinfo {author}
  {\bibfnamefont {A.~A.}\ \bibnamefont {Hyman}}, \ and\ \bibinfo {author}
  {\bibfnamefont {S.~W.}\ \bibnamefont {Grill}},\ }\href@noop {} {\bibfield
  {journal} {\bibinfo  {journal} {Science}\ }\textbf {\bibinfo {volume}
  {334}},\ \bibinfo {pages} {1137} (\bibinfo {year}
  {2011}{\natexlab{a}})}\BibitemShut {NoStop}%
\bibitem [{\citenamefont {Jilkine}(2009)}]{jilkine_2009_wave}%
  \BibitemOpen
  \bibfield  {author} {\bibinfo {author} {\bibfnamefont {A.}~\bibnamefont
  {Jilkine}},\ }\emph {\bibinfo {title} {A wave-pinning mechanism for
  eukaryotic cell polarization based on Rho GTPase dynamics}},\ \href@noop {}
  {\bibinfo {type} {{PhD} {Thesis}}},\ \bibinfo  {school} {University of
  British Columbia (Vancouver)} (\bibinfo {year} {2009})\BibitemShut {NoStop}%
\bibitem [{\citenamefont {Vanderlei}\ \emph {et~al.}(2011)\citenamefont
  {Vanderlei}, \citenamefont {Feng},\ and\ \citenamefont
  {Edelstein-Keshet}}]{vanderlei_2011_computational}%
  \BibitemOpen
  \bibfield  {author} {\bibinfo {author} {\bibfnamefont {B.}~\bibnamefont
  {Vanderlei}}, \bibinfo {author} {\bibfnamefont {J.~J.}\ \bibnamefont {Feng}},
  \ and\ \bibinfo {author} {\bibfnamefont {L.}~\bibnamefont
  {Edelstein-Keshet}},\ }\href@noop {} {\bibfield  {journal} {\bibinfo
  {journal} {Multiscale Modeling \& Simulation}\ }\textbf {\bibinfo {volume}
  {9}},\ \bibinfo {pages} {1420} (\bibinfo {year} {2011})}\BibitemShut
  {NoStop}%
\bibitem [{\citenamefont {Mori}\ \emph {et~al.}(2011)\citenamefont {Mori},
  \citenamefont {Jilkine},\ and\ \citenamefont
  {Edelstein-Keshet}}]{mori2011asymptotic}%
  \BibitemOpen
  \bibfield  {author} {\bibinfo {author} {\bibfnamefont {Y.}~\bibnamefont
  {Mori}}, \bibinfo {author} {\bibfnamefont {A.}~\bibnamefont {Jilkine}}, \
  and\ \bibinfo {author} {\bibfnamefont {L.}~\bibnamefont {Edelstein-Keshet}},\
  }\href@noop {} {\bibfield  {journal} {\bibinfo  {journal} {SIAM Journal on
  Applied Mathematics}\ }\textbf {\bibinfo {volume} {71}},\ \bibinfo {pages}
  {1401} (\bibinfo {year} {2011})}\BibitemShut {NoStop}%
\bibitem [{\citenamefont {Mar{\'e}e}\ \emph {et~al.}(2012)\citenamefont
  {Mar{\'e}e}, \citenamefont {Grieneisen},\ and\ \citenamefont
  {Edelstein-Keshet}}]{maree2012cells}%
  \BibitemOpen
  \bibfield  {author} {\bibinfo {author} {\bibfnamefont {A.~F.}\ \bibnamefont
  {Mar{\'e}e}}, \bibinfo {author} {\bibfnamefont {V.~A.}\ \bibnamefont
  {Grieneisen}}, \ and\ \bibinfo {author} {\bibfnamefont {L.}~\bibnamefont
  {Edelstein-Keshet}},\ }\href@noop {} {\bibfield  {journal} {\bibinfo
  {journal} {PLoS Computational Biology}\ }\textbf {\bibinfo {volume} {8}},\
  \bibinfo {pages} {e1002402} (\bibinfo {year} {2012})}\BibitemShut {NoStop}%
\bibitem [{\citenamefont {Mar{\'e}e}\ \emph {et~al.}(2006)\citenamefont
  {Mar{\'e}e}, \citenamefont {Jilkine}, \citenamefont {Dawes}, \citenamefont
  {Grieneisen},\ and\ \citenamefont
  {Edelstein-Keshet}}]{maree2006polarization}%
  \BibitemOpen
  \bibfield  {author} {\bibinfo {author} {\bibfnamefont {A.~F.}\ \bibnamefont
  {Mar{\'e}e}}, \bibinfo {author} {\bibfnamefont {A.}~\bibnamefont {Jilkine}},
  \bibinfo {author} {\bibfnamefont {A.}~\bibnamefont {Dawes}}, \bibinfo
  {author} {\bibfnamefont {V.~A.}\ \bibnamefont {Grieneisen}}, \ and\ \bibinfo
  {author} {\bibfnamefont {L.}~\bibnamefont {Edelstein-Keshet}},\ }\href@noop
  {} {\bibfield  {journal} {\bibinfo  {journal} {Bulletin of Mathematical
  Biology}\ }\textbf {\bibinfo {volume} {68}},\ \bibinfo {pages} {1169}
  (\bibinfo {year} {2006})}\BibitemShut {NoStop}%
\bibitem [{\citenamefont {Orlandini}\ \emph {et~al.}(2013)\citenamefont
  {Orlandini}, \citenamefont {Marenduzzo},\ and\ \citenamefont
  {Goryachev}}]{orlandini_2013_domain}%
  \BibitemOpen
  \bibfield  {author} {\bibinfo {author} {\bibfnamefont {E.}~\bibnamefont
  {Orlandini}}, \bibinfo {author} {\bibfnamefont {D.}~\bibnamefont
  {Marenduzzo}}, \ and\ \bibinfo {author} {\bibfnamefont {A.}~\bibnamefont
  {Goryachev}},\ }\href {\doibase 10.1039/C3SM50650A} {\bibfield  {journal}
  {\bibinfo  {journal} {Soft Matter}\ }\textbf {\bibinfo {volume} {9}},\
  \bibinfo {pages} {9311} (\bibinfo {year} {2013})}\BibitemShut {NoStop}%
\bibitem [{\citenamefont {Semplice}\ \emph {et~al.}(2012)\citenamefont
  {Semplice}, \citenamefont {Veglio}, \citenamefont {Naldi}, \citenamefont
  {Serini},\ and\ \citenamefont {Gamba}}]{semplice2012bistable}%
  \BibitemOpen
  \bibfield  {author} {\bibinfo {author} {\bibfnamefont {M.}~\bibnamefont
  {Semplice}}, \bibinfo {author} {\bibfnamefont {A.}~\bibnamefont {Veglio}},
  \bibinfo {author} {\bibfnamefont {G.}~\bibnamefont {Naldi}}, \bibinfo
  {author} {\bibfnamefont {G.}~\bibnamefont {Serini}}, \ and\ \bibinfo {author}
  {\bibfnamefont {A.}~\bibnamefont {Gamba}},\ }\href@noop {} {\bibfield
  {journal} {\bibinfo  {journal} {PloS ONE}\ }\textbf {\bibinfo {volume} {7}},\
  \bibinfo {pages} {e30977} (\bibinfo {year} {2012})}\BibitemShut {NoStop}%
\bibitem [{\citenamefont {Hohenberg}\ and\ \citenamefont
  {Halperin}(1977)}]{hohenberg1977theory}%
  \BibitemOpen
  \bibfield  {author} {\bibinfo {author} {\bibfnamefont {P.~C.}\ \bibnamefont
  {Hohenberg}}\ and\ \bibinfo {author} {\bibfnamefont {B.~I.}\ \bibnamefont
  {Halperin}},\ }\href@noop {} {\bibfield  {journal} {\bibinfo  {journal}
  {Reviews of Modern Physics}\ }\textbf {\bibinfo {volume} {49}},\ \bibinfo
  {pages} {435} (\bibinfo {year} {1977})}\BibitemShut {NoStop}%
\bibitem [{\citenamefont {Camley}\ and\ \citenamefont
  {Brown}(2010)}]{camley2010dynamic}%
  \BibitemOpen
  \bibfield  {author} {\bibinfo {author} {\bibfnamefont {B.~A.}\ \bibnamefont
  {Camley}}\ and\ \bibinfo {author} {\bibfnamefont {F.~L.}\ \bibnamefont
  {Brown}},\ }\href@noop {} {\bibfield  {journal} {\bibinfo  {journal}
  {Physical Review Letters}\ }\textbf {\bibinfo {volume} {105}},\ \bibinfo
  {pages} {148102} (\bibinfo {year} {2010})}\BibitemShut {NoStop}%
\bibitem [{\citenamefont {Hoege}\ and\ \citenamefont
  {Hyman}(2013)}]{hoege_2013_principles}%
  \BibitemOpen
  \bibfield  {author} {\bibinfo {author} {\bibfnamefont {C.}~\bibnamefont
  {Hoege}}\ and\ \bibinfo {author} {\bibfnamefont {A.~A.}\ \bibnamefont
  {Hyman}},\ }\href@noop {} {\bibfield  {journal} {\bibinfo  {journal} {Nature
  reviews Molecular cell biology}\ }\textbf {\bibinfo {volume} {14}},\ \bibinfo
  {pages} {315} (\bibinfo {year} {2013})}\BibitemShut {NoStop}%
\bibitem [{\citenamefont {Goehring}\ \emph
  {et~al.}(2011{\natexlab{b}})\citenamefont {Goehring}, \citenamefont {Trong},
  \citenamefont {Bois}, \citenamefont {Chowdhury}, \citenamefont {Nicola},
  \citenamefont {Hyman},\ and\ \citenamefont
  {Grill}}]{goehring_2011_polarization}%
  \BibitemOpen
  \bibfield  {author} {\bibinfo {author} {\bibfnamefont {N.~W.}\ \bibnamefont
  {Goehring}}, \bibinfo {author} {\bibfnamefont {P.~K.}\ \bibnamefont {Trong}},
  \bibinfo {author} {\bibfnamefont {J.~S.}\ \bibnamefont {Bois}}, \bibinfo
  {author} {\bibfnamefont {D.}~\bibnamefont {Chowdhury}}, \bibinfo {author}
  {\bibfnamefont {E.~M.}\ \bibnamefont {Nicola}}, \bibinfo {author}
  {\bibfnamefont {A.~A.}\ \bibnamefont {Hyman}}, \ and\ \bibinfo {author}
  {\bibfnamefont {S.~W.}\ \bibnamefont {Grill}},\ }\href {\doibase
  10.1126/science.1208619} {\bibfield  {journal} {\bibinfo  {journal}
  {Science}\ }\textbf {\bibinfo {volume} {334}},\ \bibinfo {pages} {1137}
  (\bibinfo {year} {2011}{\natexlab{b}})}\BibitemShut {NoStop}%
\bibitem [{\citenamefont {Trogdon}\ \emph {et~al.}(2018)\citenamefont
  {Trogdon}, \citenamefont {Drawert}, \citenamefont {Gomez}, \citenamefont
  {Banavar}, \citenamefont {Yi}, \citenamefont {Camp{\`a}s},\ and\
  \citenamefont {Petzold}}]{trogdon2018effect}%
  \BibitemOpen
  \bibfield  {author} {\bibinfo {author} {\bibfnamefont {M.}~\bibnamefont
  {Trogdon}}, \bibinfo {author} {\bibfnamefont {B.}~\bibnamefont {Drawert}},
  \bibinfo {author} {\bibfnamefont {C.}~\bibnamefont {Gomez}}, \bibinfo
  {author} {\bibfnamefont {S.~P.}\ \bibnamefont {Banavar}}, \bibinfo {author}
  {\bibfnamefont {T.-M.}\ \bibnamefont {Yi}}, \bibinfo {author} {\bibfnamefont
  {O.}~\bibnamefont {Camp{\`a}s}}, \ and\ \bibinfo {author} {\bibfnamefont
  {L.~R.}\ \bibnamefont {Petzold}},\ }\href@noop {} {\bibfield  {journal}
  {\bibinfo  {journal} {PLoS Computational Biology}\ }\textbf {\bibinfo
  {volume} {14}},\ \bibinfo {pages} {e1006241} (\bibinfo {year}
  {2018})}\BibitemShut {NoStop}%
\bibitem [{\citenamefont {Klinkert}\ \emph {et~al.}(2019)\citenamefont
  {Klinkert}, \citenamefont {Levernier}, \citenamefont {Gross}, \citenamefont
  {Gentili}, \citenamefont {von Tobel}, \citenamefont {Pierron}, \citenamefont
  {Busso}, \citenamefont {Herrman}, \citenamefont {Grill}, \citenamefont
  {Kruse},\ and\ \citenamefont {Gönczy}}]{klinkert_2019_aurora}%
  \BibitemOpen
  \bibfield  {author} {\bibinfo {author} {\bibfnamefont {K.}~\bibnamefont
  {Klinkert}}, \bibinfo {author} {\bibfnamefont {N.}~\bibnamefont {Levernier}},
  \bibinfo {author} {\bibfnamefont {P.}~\bibnamefont {Gross}}, \bibinfo
  {author} {\bibfnamefont {C.}~\bibnamefont {Gentili}}, \bibinfo {author}
  {\bibfnamefont {L.}~\bibnamefont {von Tobel}}, \bibinfo {author}
  {\bibfnamefont {M.}~\bibnamefont {Pierron}}, \bibinfo {author} {\bibfnamefont
  {C.}~\bibnamefont {Busso}}, \bibinfo {author} {\bibfnamefont
  {S.}~\bibnamefont {Herrman}}, \bibinfo {author} {\bibfnamefont {S.~W.}\
  \bibnamefont {Grill}}, \bibinfo {author} {\bibfnamefont {K.}~\bibnamefont
  {Kruse}}, \ and\ \bibinfo {author} {\bibfnamefont {P.}~\bibnamefont
  {Gönczy}},\ }\href {\doibase 10.7554/eLife.44552} {\bibfield  {journal}
  {\bibinfo  {journal} {eLife}\ }\textbf {\bibinfo {volume} {8}},\ \bibinfo
  {pages} {e44552} (\bibinfo {year} {2019})}\BibitemShut {NoStop}%
\bibitem [{\citenamefont {Koo}\ \emph {et~al.}(2015)\citenamefont {Koo},
  \citenamefont {Weitzman}, \citenamefont {Sabanaygam}, \citenamefont {van
  Golen},\ and\ \citenamefont {Mochrie}}]{Koo_2015}%
  \BibitemOpen
  \bibfield  {author} {\bibinfo {author} {\bibfnamefont {P.~K.}\ \bibnamefont
  {Koo}}, \bibinfo {author} {\bibfnamefont {M.}~\bibnamefont {Weitzman}},
  \bibinfo {author} {\bibfnamefont {C.~R.}\ \bibnamefont {Sabanaygam}},
  \bibinfo {author} {\bibfnamefont {K.~L.}\ \bibnamefont {van Golen}}, \ and\
  \bibinfo {author} {\bibfnamefont {S.~G.~J.}\ \bibnamefont {Mochrie}},\ }\href
  {\doibase 10.1371/journal.pcbi.1004297} {\bibfield  {journal} {\bibinfo
  {journal} {{PLOS} Computational Biology}\ }\textbf {\bibinfo {volume} {11}},\
  \bibinfo {pages} {e1004297} (\bibinfo {year} {2015})}\BibitemShut {NoStop}%
\bibitem [{\citenamefont {Goryachev}\ and\ \citenamefont
  {Pokhilko}(2008)}]{goryachev2008dynamics}%
  \BibitemOpen
  \bibfield  {author} {\bibinfo {author} {\bibfnamefont {A.~B.}\ \bibnamefont
  {Goryachev}}\ and\ \bibinfo {author} {\bibfnamefont {A.~V.}\ \bibnamefont
  {Pokhilko}},\ }\href@noop {} {\bibfield  {journal} {\bibinfo  {journal} {FEBS
  Letters}\ }\textbf {\bibinfo {volume} {582}},\ \bibinfo {pages} {1437}
  (\bibinfo {year} {2008})}\BibitemShut {NoStop}%
\bibitem [{\citenamefont {Tamemoto}\ and\ \citenamefont
  {Noguchi}(2020)}]{tamemoto_2020_pattern}%
  \BibitemOpen
  \bibfield  {author} {\bibinfo {author} {\bibfnamefont {N.}~\bibnamefont
  {Tamemoto}}\ and\ \bibinfo {author} {\bibfnamefont {H.}~\bibnamefont
  {Noguchi}},\ }\href {\doibase 10.1038/s41598-020-76695-x} {\bibfield
  {journal} {\bibinfo  {journal} {Scientific Reports}\ }\textbf {\bibinfo
  {volume} {10}} (\bibinfo {year} {2020}),\
  10.1038/s41598-020-76695-x}\BibitemShut {NoStop}%
\bibitem [{\citenamefont {Tamemoto}\ and\ \citenamefont
  {Noguchi}(2021)}]{tamemoto_2021_reaction}%
  \BibitemOpen
  \bibfield  {author} {\bibinfo {author} {\bibfnamefont {N.}~\bibnamefont
  {Tamemoto}}\ and\ \bibinfo {author} {\bibfnamefont {H.}~\bibnamefont
  {Noguchi}},\ }\href@noop {} {\bibfield  {journal} {\bibinfo  {journal} {Soft
  Matter}\ }\textbf {\bibinfo {volume} {17}},\ \bibinfo {pages} {6589}
  (\bibinfo {year} {2021})}\BibitemShut {NoStop}%
\bibitem [{\citenamefont {Tateno}\ and\ \citenamefont
  {Ishihara}(2021)}]{tateno2021interfacial}%
  \BibitemOpen
  \bibfield  {author} {\bibinfo {author} {\bibfnamefont {M.}~\bibnamefont
  {Tateno}}\ and\ \bibinfo {author} {\bibfnamefont {S.}~\bibnamefont
  {Ishihara}},\ }\href@noop {} {\bibfield  {journal} {\bibinfo  {journal}
  {Physical Review Research}\ }\textbf {\bibinfo {volume} {3}},\ \bibinfo
  {pages} {023198} (\bibinfo {year} {2021})}\BibitemShut {NoStop}%
\bibitem [{\citenamefont {Bergmann}\ \emph {et~al.}(2018)\citenamefont
  {Bergmann}, \citenamefont {Rapp},\ and\ \citenamefont
  {Zimmermann}}]{bergmann2018active}%
  \BibitemOpen
  \bibfield  {author} {\bibinfo {author} {\bibfnamefont {F.}~\bibnamefont
  {Bergmann}}, \bibinfo {author} {\bibfnamefont {L.}~\bibnamefont {Rapp}}, \
  and\ \bibinfo {author} {\bibfnamefont {W.}~\bibnamefont {Zimmermann}},\
  }\href@noop {} {\bibfield  {journal} {\bibinfo  {journal} {Physical Review
  E}\ }\textbf {\bibinfo {volume} {98}},\ \bibinfo {pages} {020603} (\bibinfo
  {year} {2018})}\BibitemShut {NoStop}%
\bibitem [{\citenamefont {Brauns}\ \emph {et~al.}(2020)\citenamefont {Brauns},
  \citenamefont {Halatek},\ and\ \citenamefont {Frey}}]{brauns2020phase}%
  \BibitemOpen
  \bibfield  {author} {\bibinfo {author} {\bibfnamefont {F.}~\bibnamefont
  {Brauns}}, \bibinfo {author} {\bibfnamefont {J.}~\bibnamefont {Halatek}}, \
  and\ \bibinfo {author} {\bibfnamefont {E.}~\bibnamefont {Frey}},\ }\href@noop
  {} {\bibfield  {journal} {\bibinfo  {journal} {Physical Review X}\ }\textbf
  {\bibinfo {volume} {10}},\ \bibinfo {pages} {041036} (\bibinfo {year}
  {2020})}\BibitemShut {NoStop}%
\bibitem [{\citenamefont {Ge{\ss}ele}\ \emph {et~al.}(2020)\citenamefont
  {Ge{\ss}ele}, \citenamefont {Halatek}, \citenamefont {W{\"u}rthner},\ and\
  \citenamefont {Frey}}]{gessele2020geometric}%
  \BibitemOpen
  \bibfield  {author} {\bibinfo {author} {\bibfnamefont {R.}~\bibnamefont
  {Ge{\ss}ele}}, \bibinfo {author} {\bibfnamefont {J.}~\bibnamefont {Halatek}},
  \bibinfo {author} {\bibfnamefont {L.}~\bibnamefont {W{\"u}rthner}}, \ and\
  \bibinfo {author} {\bibfnamefont {E.}~\bibnamefont {Frey}},\ }\href@noop {}
  {\bibfield  {journal} {\bibinfo  {journal} {Nature Communications}\ }\textbf
  {\bibinfo {volume} {11}},\ \bibinfo {pages} {1} (\bibinfo {year}
  {2020})}\BibitemShut {NoStop}%
\bibitem [{\citenamefont {Cannon}\ \emph {et~al.}(2017)\citenamefont {Cannon},
  \citenamefont {Woods},\ and\ \citenamefont
  {Gladfelter}}]{cannon_2017_unsolved}%
  \BibitemOpen
  \bibfield  {author} {\bibinfo {author} {\bibfnamefont {K.~S.}\ \bibnamefont
  {Cannon}}, \bibinfo {author} {\bibfnamefont {B.~L.}\ \bibnamefont {Woods}}, \
  and\ \bibinfo {author} {\bibfnamefont {A.~S.}\ \bibnamefont {Gladfelter}},\
  }\href {\doibase 10.1016/j.tibs.2017.10.001} {\bibfield  {journal} {\bibinfo
  {journal} {Trends in Biochemical Sciences}\ }\textbf {\bibinfo {volume}
  {42}},\ \bibinfo {pages} {961} (\bibinfo {year} {2017})}\BibitemShut
  {NoStop}%
\bibitem [{\citenamefont {Cannon}\ \emph {et~al.}(2019)\citenamefont {Cannon},
  \citenamefont {Woods}, \citenamefont {Crutchley},\ and\ \citenamefont
  {Gladfelter}}]{cannon_2019_amphipathic}%
  \BibitemOpen
  \bibfield  {author} {\bibinfo {author} {\bibfnamefont {K.~S.}\ \bibnamefont
  {Cannon}}, \bibinfo {author} {\bibfnamefont {B.~L.}\ \bibnamefont {Woods}},
  \bibinfo {author} {\bibfnamefont {J.~M.}\ \bibnamefont {Crutchley}}, \ and\
  \bibinfo {author} {\bibfnamefont {A.~S.}\ \bibnamefont {Gladfelter}},\ }\href
  {\doibase 10.1083/jcb.201807211} {\bibfield  {journal} {\bibinfo  {journal}
  {The Journal of Cell Biology}\ }\textbf {\bibinfo {volume} {218}},\ \bibinfo
  {pages} {1128} (\bibinfo {year} {2019})}\BibitemShut {NoStop}%
\bibitem [{\citenamefont {Gowrishankar}\ \emph {et~al.}(2012)\citenamefont
  {Gowrishankar}, \citenamefont {Ghosh}, \citenamefont {Saha}, \citenamefont
  {Rumamol}, \citenamefont {Mayor},\ and\ \citenamefont
  {Rao}}]{gowrishankar2012active}%
  \BibitemOpen
  \bibfield  {author} {\bibinfo {author} {\bibfnamefont {K.}~\bibnamefont
  {Gowrishankar}}, \bibinfo {author} {\bibfnamefont {S.}~\bibnamefont {Ghosh}},
  \bibinfo {author} {\bibfnamefont {S.}~\bibnamefont {Saha}}, \bibinfo {author}
  {\bibfnamefont {C.}~\bibnamefont {Rumamol}}, \bibinfo {author} {\bibfnamefont
  {S.}~\bibnamefont {Mayor}}, \ and\ \bibinfo {author} {\bibfnamefont
  {M.}~\bibnamefont {Rao}},\ }\href@noop {} {\bibfield  {journal} {\bibinfo
  {journal} {Cell}\ }\textbf {\bibinfo {volume} {149}},\ \bibinfo {pages}
  {1353} (\bibinfo {year} {2012})}\BibitemShut {NoStop}%
\bibitem [{\citenamefont {Kusumi}\ \emph {et~al.}(2011)\citenamefont {Kusumi},
  \citenamefont {Suzuki}, \citenamefont {Kasai}, \citenamefont {Ritchie},\ and\
  \citenamefont {Fujiwara}}]{kusumi2011hierarchical}%
  \BibitemOpen
  \bibfield  {author} {\bibinfo {author} {\bibfnamefont {A.}~\bibnamefont
  {Kusumi}}, \bibinfo {author} {\bibfnamefont {K.~G.}\ \bibnamefont {Suzuki}},
  \bibinfo {author} {\bibfnamefont {R.~S.}\ \bibnamefont {Kasai}}, \bibinfo
  {author} {\bibfnamefont {K.}~\bibnamefont {Ritchie}}, \ and\ \bibinfo
  {author} {\bibfnamefont {T.~K.}\ \bibnamefont {Fujiwara}},\ }\href@noop {}
  {\bibfield  {journal} {\bibinfo  {journal} {Trends in Biochemical Sciences}\
  }\textbf {\bibinfo {volume} {36}},\ \bibinfo {pages} {604} (\bibinfo {year}
  {2011})}\BibitemShut {NoStop}%
\bibitem [{\citenamefont {Hosaka}\ \emph {et~al.}(2017)\citenamefont {Hosaka},
  \citenamefont {Yasuda}, \citenamefont {Okamoto},\ and\ \citenamefont
  {Komura}}]{hosaka2017lateral}%
  \BibitemOpen
  \bibfield  {author} {\bibinfo {author} {\bibfnamefont {Y.}~\bibnamefont
  {Hosaka}}, \bibinfo {author} {\bibfnamefont {K.}~\bibnamefont {Yasuda}},
  \bibinfo {author} {\bibfnamefont {R.}~\bibnamefont {Okamoto}}, \ and\
  \bibinfo {author} {\bibfnamefont {S.}~\bibnamefont {Komura}},\ }\href@noop {}
  {\bibfield  {journal} {\bibinfo  {journal} {Physical Review E}\ }\textbf
  {\bibinfo {volume} {95}},\ \bibinfo {pages} {052407} (\bibinfo {year}
  {2017})}\BibitemShut {NoStop}%
\bibitem [{\citenamefont {Swartz}\ and\ \citenamefont
  {Camley}(2021)}]{swartz2021active}%
  \BibitemOpen
  \bibfield  {author} {\bibinfo {author} {\bibfnamefont {D.~W.}\ \bibnamefont
  {Swartz}}\ and\ \bibinfo {author} {\bibfnamefont {B.~A.}\ \bibnamefont
  {Camley}},\ }\href@noop {} {\bibfield  {journal} {\bibinfo  {journal} {Soft
  Matter}\ }\textbf {\bibinfo {volume} {17}},\ \bibinfo {pages} {9876}
  (\bibinfo {year} {2021})}\BibitemShut {NoStop}%
\bibitem [{\citenamefont {Edelstein-Keshet}\ \emph {et~al.}(2013)\citenamefont
  {Edelstein-Keshet}, \citenamefont {Holmes}, \citenamefont {Zajac},\ and\
  \citenamefont {Dutot}}]{edelstein2013simple}%
  \BibitemOpen
  \bibfield  {author} {\bibinfo {author} {\bibfnamefont {L.}~\bibnamefont
  {Edelstein-Keshet}}, \bibinfo {author} {\bibfnamefont {W.~R.}\ \bibnamefont
  {Holmes}}, \bibinfo {author} {\bibfnamefont {M.}~\bibnamefont {Zajac}}, \
  and\ \bibinfo {author} {\bibfnamefont {M.}~\bibnamefont {Dutot}},\
  }\href@noop {} {\bibfield  {journal} {\bibinfo  {journal} {Philosophical
  Transactions of the Royal Society B: Biological Sciences}\ }\textbf {\bibinfo
  {volume} {368}},\ \bibinfo {pages} {20130003} (\bibinfo {year}
  {2013})}\BibitemShut {NoStop}%
\bibitem [{\citenamefont {Mogilner}\ \emph {et~al.}(2012)\citenamefont
  {Mogilner}, \citenamefont {Allard},\ and\ \citenamefont
  {Wollman}}]{mogilner2012cell}%
  \BibitemOpen
  \bibfield  {author} {\bibinfo {author} {\bibfnamefont {A.}~\bibnamefont
  {Mogilner}}, \bibinfo {author} {\bibfnamefont {J.}~\bibnamefont {Allard}}, \
  and\ \bibinfo {author} {\bibfnamefont {R.}~\bibnamefont {Wollman}},\
  }\href@noop {} {\bibfield  {journal} {\bibinfo  {journal} {Science}\ }\textbf
  {\bibinfo {volume} {336}},\ \bibinfo {pages} {175} (\bibinfo {year}
  {2012})}\BibitemShut {NoStop}%
\bibitem [{\citenamefont {Holmes}\ \emph {et~al.}(2017)\citenamefont {Holmes},
  \citenamefont {Park}, \citenamefont {Levchenko},\ and\ \citenamefont
  {Edelstein-Keshet}}]{holmes2017mathematical}%
  \BibitemOpen
  \bibfield  {author} {\bibinfo {author} {\bibfnamefont {W.~R.}\ \bibnamefont
  {Holmes}}, \bibinfo {author} {\bibfnamefont {J.}~\bibnamefont {Park}},
  \bibinfo {author} {\bibfnamefont {A.}~\bibnamefont {Levchenko}}, \ and\
  \bibinfo {author} {\bibfnamefont {L.}~\bibnamefont {Edelstein-Keshet}},\
  }\href@noop {} {\bibfield  {journal} {\bibinfo  {journal} {PLoS computational
  biology}\ }\textbf {\bibinfo {volume} {13}},\ \bibinfo {pages} {e1005524}
  (\bibinfo {year} {2017})}\BibitemShut {NoStop}%
\bibitem [{\citenamefont {Rognes}\ \emph {et~al.}(2013)\citenamefont {Rognes},
  \citenamefont {Ham}, \citenamefont {Cotter},\ and\ \citenamefont
  {McRae}}]{rognes_2013_automating}%
  \BibitemOpen
  \bibfield  {author} {\bibinfo {author} {\bibfnamefont {M.~E.}\ \bibnamefont
  {Rognes}}, \bibinfo {author} {\bibfnamefont {D.~A.}\ \bibnamefont {Ham}},
  \bibinfo {author} {\bibfnamefont {C.~J.}\ \bibnamefont {Cotter}}, \ and\
  \bibinfo {author} {\bibfnamefont {A.~T.~T.}\ \bibnamefont {McRae}},\ }\href
  {\doibase 10.5194/gmd-6-2099-2013} {\bibfield  {journal} {\bibinfo  {journal}
  {Geoscientific Model Development}\ }\textbf {\bibinfo {volume} {6}},\
  \bibinfo {pages} {2099} (\bibinfo {year} {2013})}\BibitemShut {NoStop}%
\bibitem [{\citenamefont {Aln{\ae}s}\ \emph {et~al.}(2015)\citenamefont
  {Aln{\ae}s}, \citenamefont {Blechta}, \citenamefont {Hake}, \citenamefont
  {Johansson}, \citenamefont {Kehlet}, \citenamefont {Logg}, \citenamefont
  {Richardson}, \citenamefont {Ring}, \citenamefont {Rognes},\ and\
  \citenamefont {Wells}}]{alnaes2015fenics}%
  \BibitemOpen
  \bibfield  {author} {\bibinfo {author} {\bibfnamefont {M.}~\bibnamefont
  {Aln{\ae}s}}, \bibinfo {author} {\bibfnamefont {J.}~\bibnamefont {Blechta}},
  \bibinfo {author} {\bibfnamefont {J.}~\bibnamefont {Hake}}, \bibinfo {author}
  {\bibfnamefont {A.}~\bibnamefont {Johansson}}, \bibinfo {author}
  {\bibfnamefont {B.}~\bibnamefont {Kehlet}}, \bibinfo {author} {\bibfnamefont
  {A.}~\bibnamefont {Logg}}, \bibinfo {author} {\bibfnamefont {C.}~\bibnamefont
  {Richardson}}, \bibinfo {author} {\bibfnamefont {J.}~\bibnamefont {Ring}},
  \bibinfo {author} {\bibfnamefont {M.~E.}\ \bibnamefont {Rognes}}, \ and\
  \bibinfo {author} {\bibfnamefont {G.~N.}\ \bibnamefont {Wells}},\ }\href@noop
  {} {\bibfield  {journal} {\bibinfo  {journal} {Archive of Numerical
  Software}\ }\textbf {\bibinfo {volume} {3}} (\bibinfo {year}
  {2015})}\BibitemShut {NoStop}%
\end{thebibliography}
\end{document}